\documentclass[aps,prd,reprint,showpacs,superscriptaddress,twocolumn]{revtex4-2}
\usepackage{amsfonts}
\usepackage{amsmath}
\usepackage{amssymb}
\usepackage{dsfont}
\usepackage{hyperref}
\usepackage{graphicx}
\usepackage{float}
\usepackage[usenames,dvipsnames]{xcolor}
\usepackage[normalem]{ulem}
\usepackage{times}
\usepackage{braket}
\usepackage{booktabs}
\usepackage{caption}
\usepackage{subcaption}

\usepackage{natbib}
\usepackage{physics}
\usepackage{mathtools}

\newcommand{\orcid}[1]{\href{https://orcid.org/#1}
	{\includegraphics[width=7pt]{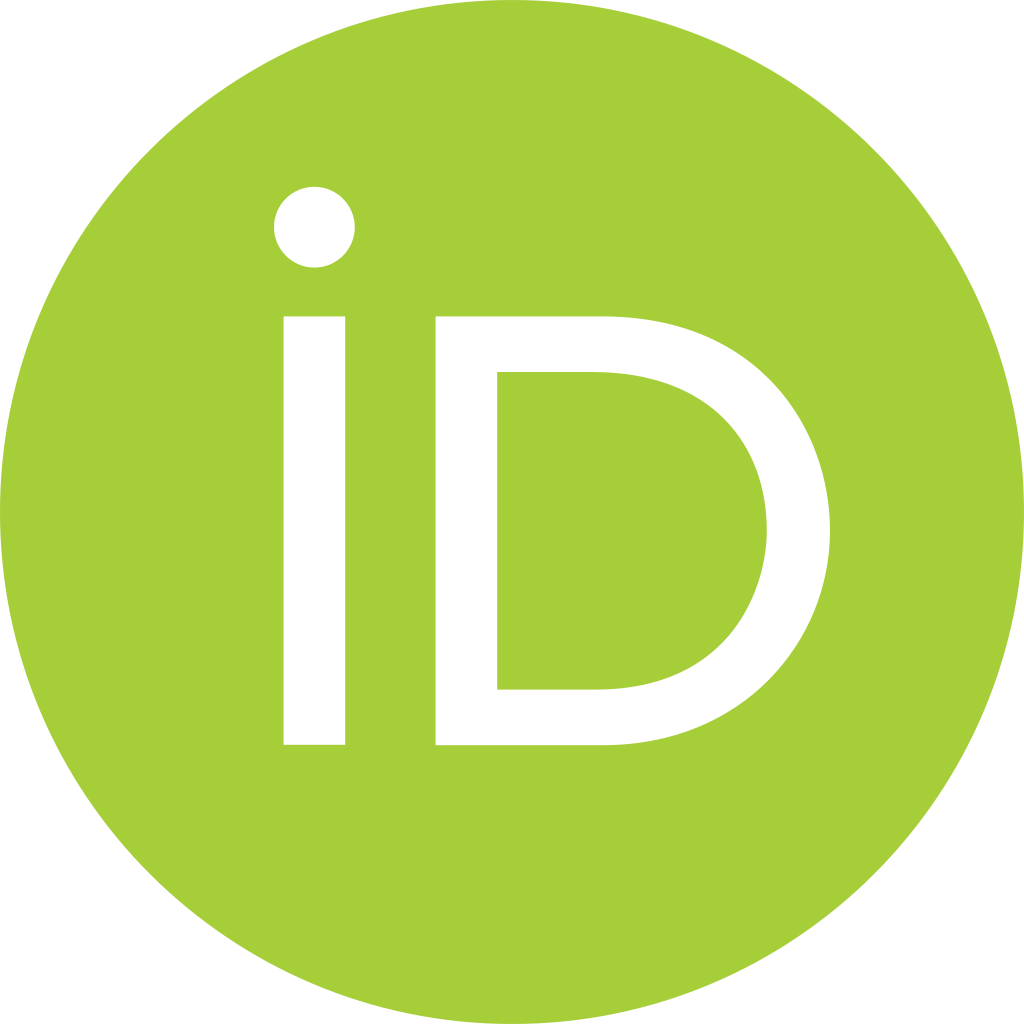}}}

\hypersetup{
    colorlinks=true,linkcolor=blue,citecolor=blue,
    filecolor=blue,urlcolor=blue,breaklinks=true
}
\RequirePackage{color}

\begin{document}

\date{\today}

\title{Study on the effects of anisotropic effective mass on electronic properties, magnetization and persistent current in semiconductor quantum ring with conical geometry}

\author{Francisco A. G. de Lira}
\email{francisco.augoncalves@gmail.com}
\affiliation{Departamento de F\'{\i}sica, Universidade Federal do Maranh\~{a}o, 65085-580 S\~{a}o Lu\'{\i}s, Maranh\~{a}o, Brazil}

\author{Lu\'{i}s Fernando C. Pereira}
\email{luisfernandofisica@hotmail.com}
\affiliation{Departamento de F\'{\i}sica, Universidade Federal do Maranh\~{a}o, 65085-580 S\~{a}o Lu\'{\i}s, Maranh\~{a}o, Brazil}

\author{Edilberto O. Silva \orcid{0000-0002-0297-5747}}
\email{edilberto.silva@ufma.br}
\affiliation{Departamento de F\'{\i}sica, Universidade Federal do Maranh\~{a}o, 65085-580 S\~{a}o Lu\'{\i}s, Maranh\~{a}o, Brazil}

\begin{abstract}
We study a 2D mesoscopic ring with an anisotropic effective mass considering surface quantum confinement effects. Consider that the ring is defined on the surface of a cone, which can be controlled topologically and mapped to the 2D ring in flat space. We demonstrate through numerical analysis that the electronic properties, the magnetization, and the persistent current undergo significant changes due to quantum confinement and non-isotropic mass. We investigate these changes in the direct band gap semiconductors SiC, ZnO, GaN, and AlN. There is a plus (or minus) shift in the energy sub-bands for different values of curvature parameter and anisotropy. Manifestations of this nature are also seen in the Fermi energy profile as a function of the magnetic field and in the ring width as a function of the curvature parameter. Aharonov-Bohm (AB) and de Haas van-Alphen (dHvA) oscillations are also studied, and we find that they are sensitive to variations in curvature and anisotropy. 
\end{abstract}

\maketitle

\section{Introduction}\label{intro}

In recent years, there has been a surge in research focused on elucidating the intricate properties of mesoscopic materials, encompassing their optical, electronic, magnetic, and transport characteristics. This burgeoning interest stems from the diverse spectrum of applications that these materials offer, ranging from the engineering of electronic circuits to transformative breakthroughs in the field of medicine \cite{dasgupta2010new,buzki2016micro, JIM.2021.290.486}. The exploration of mesoscopic materials has unveiled a multitude of intriguing phenomena and potential technological advancements. Researchers have been driven to delve deeper into the intricate interplay between the fundamental properties of these materials and their practical applications \cite{shevchenko2019mesoscopic}. By gaining a comprehensive understanding of their optical properties, including absorption, emission, and scattering behavior, researchers aim to exploit these materials for advanced optical devices, imaging technologies, and sensing applications \cite{sanchez2021quantum}. Furthermore, the electronic properties of mesoscopic materials have captured the attention of researchers seeking to harness their unique electronic behavior to develop novel electronic devices with enhanced functionalities. The investigation of phenomena, such as quantum confinement, carrier transport, and bandgap engineering, in these materials holds promise for the realization of ultrafast transistors, high-performance sensors, and energy-efficient devices \cite{kodolov2016applied,datta1997electronic}.

The magnetic characteristics of mesoscopic materials have also emerged as a focal point of research. By exploring phenomena such as spin dynamics, magnetic ordering, and spin-dependent transport, scientists endeavor to unlock the potential of these materials for spintronic applications, magnetic data storage, and magnetic field sensing \cite{bookRuiz}. Additionally, the transport properties of mesoscopic materials have attracted significant attention. Researchers strive to comprehend these material's charge, heat, and spin transport mechanisms. By unraveling the intricacies of carrier mobility, thermal conductivity, and spin diffusion, scientists aim to develop materials with superior transport properties for applications in energy conversion, thermoelectric devices, and spintronic circuits \cite{li2017spin}. 

Overall, the multifaceted exploration of mesoscopic materials has become a thriving field of research, driven by their wide-ranging applications and the desire to uncover their fundamental properties. Through meticulous investigation and innovative experimentation, researchers seek to unlock the full potential of these materials, propelling advancements in various technological domains \cite{lerner2004fundamental}. One highly successful example of such materials is graphene \cite{Science.2004.306.666}, a two-dimensional structure known for its exceptional thermal and electric conductivity, as well as its remarkable mechanical resistance and malleability. Consequently, graphene layers have garnered significant attention for applications in various fields, including nanoelectronics \cite{PARK201613} and drug delivery systems \cite{LIU20139243}.
	
An important mesoscopic structure is a quantum ring, which can confine particles such as highly mobile electrons within a distinct region. This confinement leads to the manifestation of important physical phenomena, such as the AB effect \cite{Bachtold1999}, dHvA oscillations, persistent currents in non-superconducting metals \cite{PhysRevLett.70.2020}, and the quantum Hall effect. While many theoretical and experimental studies on quantum rings have focused on a GaAs/AlAs semiconductor heterostructure with an isotropic electron effective mass \cite{PhysRevLett.70.2020,SST.1996.11.1635,PRB.1999.60.5626}, it is important to note that semiconductors, in general, exhibit anisotropic electron-effective masses, including materials like Si, Ge, ZnO, and SiC. Therefore, the aim of our study is to analyze the effects of anisotropic effective mass on the physical properties of quantum rings. Furthermore, we are interested in exploring the influence of conical curvature in the system since this type of geometry has already been experimentally achieved in carbon lattices \cite{PhysRevLett.115.026801,doi:10.1063/1.1377859}, and it is well-established that the presence of defects in materials, such as impurities, can induce significant changes in particular properties of these systems \cite{book.2012.Ching,addou2022defects}. 
	
The organization of our work is as follows. In Sec. \ref{sec2}, we present the theoretical model of anisotropic effective mass and the reference materials studied. In Sec. \ref{sec3}, we introduce the conical geometry representing the samples analyzed and describe their surface geometric properties. In Sec. \ref{sec4}, we solve the Schrödinger equation in conic geometry and determine the energies and wave functions. We discuss the electronic properties in Sections \ref{sec5} and \ref{sec6}. The Secs. \ref{sec7} and \ref{sec8} are devoted to studying magnetization and persistent current in the ring, respectively. In Sec. \ref{sec9}, we discuss an application in the context of spins. Conclusions are presented in Sec. \ref{sec10}.
	
\section{Anisotropic effective mass}\label{sec2}
	
The effective mass approximation in semiconductor materials is applied by replacing an electron in complex interaction with the crystal lattice for a `free' electron responding to an electric field with a mass value different from the usual, which is known to be $m_{e}\approx 9.11 \times 10^{-31}\hspace{0.05cm}\text{kg}$.
	
Direct band gap semiconductors are those which have their conduction band minimum located in the center of their Brillouin zone. Examples such as GaAs and InAs present an isotropic value of effective mass in all directions on the reciprocal space because their Brillouin zones are equidistant from the origin. However, for other direct band ones, such as SiC and ZnO, that are able to crystallize in a \textit{hexagonal wurtzite structure} \cite{ALGARNI2019102694,C9CE00746F,inbook}, anisotropy will rise from the different free electron energy on their Brillouin zone limits. The Brillouin zone symmetry for these anisotropic materials is characterized by a transverse plane and a longitudinal direction, leading to an ellipsoidal effective mass defined by two parameters: a transverse effective mass $m_{t}$ and a longitudinal effective mass $m_{l}$ \cite{kittel, semiconductors}. Indirect band gaps semiconductors such as Si and Ge show more complex anisotropy since the minimum conduction band is not in the center of reciprocal space.  
	
By assuming that the longitudinal direction coincides with the \textit{z}-axis, we have the following dispersion relation \cite{Kittel2004}:
\begin{equation}
\epsilon(\mathbf{k})=\dfrac{\hbar^{2}}{2}\left(\dfrac{k^{2}_{x}+k^{2}_{y}}{m_{t}}+\dfrac{k^{2}_{z}}{m_{l}}\right), 
\label{anis01}
\end{equation}
which, for an given value of $\epsilon(\mathbf{k})$, represents an ellipsoidal surface in reciprocal space.
	
\begin{figure}[!h]
\centering
\begin{subfigure}{0.2\textwidth}
\includegraphics[width=\textwidth]{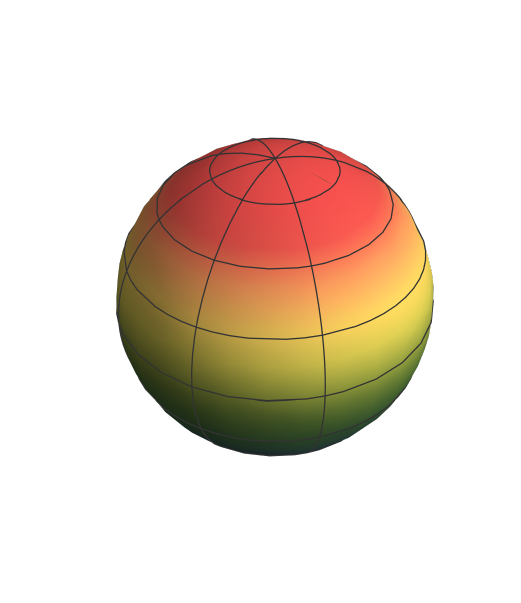}
\subcaption{$R=1.000$}
\label{elipsoidal01}
\end{subfigure}
\begin{subfigure}{0.2\textwidth}
\includegraphics[width=\textwidth]{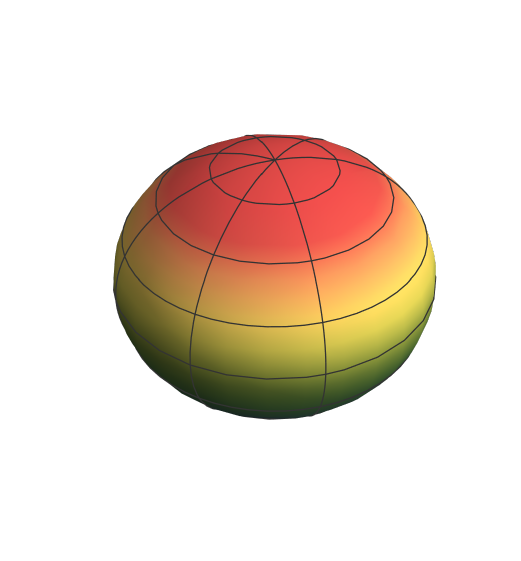}
\subcaption{$R=1.500$}
\label{elipsoidal02}
\end{subfigure}
\begin{subfigure}{0.2\textwidth}
\includegraphics[width=\textwidth]{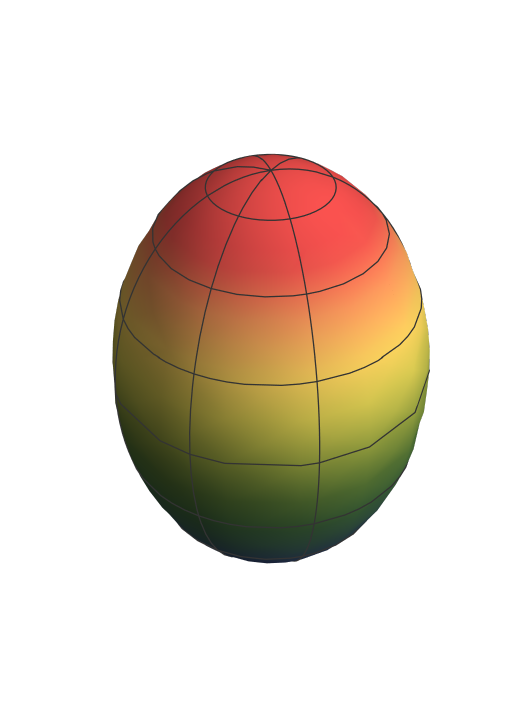}
\subcaption{$R=0.500$}
\label{elipsoidal03}
\end{subfigure}
\caption{Representation of the Fermi surfaces for different values of $R$ parameter. Spherical ($R=1$), oblate spheroid ($R>1$), prolate spheroid ($R<1$).}\label{elipsoidal}
\end{figure}

Our approach for analyzing the effects of anisotropy is based on the deformation of the spherically symmetric effective mass of the GaAs $\mu=0.067m_{e}$ into an ellipsoidal surface described by Eq. (\ref{anis01}) and illustrated in Fig. \ref{elipsoidal}. Thus, we relate the anisotropic masses with $\mu$ by $m_{t}=\sigma\mu$ and $m_{l}=\beta\mu$, where $\sigma,\beta \in\mathbb{N}^{*}$. Moreover, we define $R=m_{t}/m_{l}=\sigma/\beta$ as the ratio between these two mass values. When $R=1$ (Fig. \ref{elipsoidal01}), we have a spherically symmetric Fermi surface. On the other side, for $R>1$ and $R<1$ we obtain oblate and prolate spheroids, respectively, in Figs \ref{elipsoidal02} and \ref{elipsoidal03}.  
	
Consider now a sample of doped GaAs at zero temperature and with volume $V=L_{x}L_{y}L_{z}$ containing $N_{e}$ electrons in the conduction band provided by the impurities (extrinsic carriers). If $E_{f}$ is the Fermi energy of the system, the volume enclosed by the isotropic Fermi surface is given by
\begin{equation}
\mathcal{V}=\dfrac{4}{3}\pi\left(\dfrac{2E_{f}}{\hbar^2}\right)^{3/2}\mu^{3/2}.
\end{equation}
By considering a sample with an anisotropic effective mass of the same physical volume $V$ and using the definitions above in dispersion relation (\ref{anis01}), we get an equation of an ellipsoidal surface in reciprocal space given by
\begin{equation}
\dfrac{k^2_{x}}{a^2}+\dfrac{k^2_{y}}{b^2}+\dfrac{k^2_{z}}{c^2}=1,
\label{anis02}
\end{equation}
with the semi-axes $a=b=(2\sigma\mu E'_{f}/\hbar^2)^{1/2}$ and $c=(2\beta\mu E'_{f}/\hbar^2)^{1/2}$, where $E'_{f}$ represents the respective Fermi energy not necessarily equal to $E_{f}$. The volume enclosed by this ellipsoidal surface is
\begin{equation}
\mathcal{V'}=\dfrac{4}{3}\pi\left(\dfrac{2E'_{f}}{\hbar^2}\right)^{3/2}m_{l}^{1/2}m_{t}
\end{equation}
	
Since we want to investigate the effects of the deformation in the isotropic Fermi surface, we require that $\mathcal{V}=\mathcal{V'}$ and $E_{f}=E'_{f}$. Thus, $\sigma^2\beta=1$ and the anisotropic masses are related in terms of $\mu$ by
\begin{equation}
m_{t}=R^{1/3}\mu\quad\text{and}\quad m_{l}=R^{-2/3}\mu.
\label{anis03}
\end{equation}
A direct consequence of the constraint for both volume and Fermi energy is that the number of extrinsic carriers of both isotropic and anisotropic systems will remain constant. In fact, from Eq. (\ref{anis01}) the $\mathbf{k'}$ vector for the anisotropic system is given by
\begin{equation}
\epsilon(\mathbf{k'})=\dfrac{\hbar^2\mathbf{k'}^2}{2\mu},\qquad \mathbf{k'}=2\pi\left(\dfrac{n_{x}}{\sqrt{\sigma} L_{x}},\dfrac{n_{x}}{\sqrt{\sigma} L_{x}},\dfrac{n_{x}}{\sqrt{\beta} L_{x}}\right). \label{anis04}
\end{equation}
Then, the volume of the primitive cell $\upsilon'$  in the reciprocal space will be equal to the one of the isotropic $\upsilon$:
\begin{equation}
\upsilon'=\dfrac{8\pi^3}{\sqrt{(\sigma^{2}\beta})L_{x}L_{y}L_{z}}=\dfrac{8\pi^3}{V}=\upsilon.
\label{anis05}
\end{equation}
	
In Table \ref{tab1}, we display the values of ratio $R$ for some direct band gap semiconductors that are now able to crystallize in the wurtzite structure \cite{SHOKHOVETS2006299, doi:10.1063/1.113576, PhysRevB.52.8132}. We use these values in the next sections to obtain quantitative properties as well as to identify the effects due to different values of ellipsoidal anisotropy in the model.
\begin{table}[h]
\centering
\setlength{\tabcolsep}{12 pt}
\vspace{0.15 cm}
\begin{tabular}{cccc}
\hline
\multicolumn{1}{c}{\textbf{Semiconductor}}
& \multicolumn{1}{c}{$m_{l}$} & \multicolumn{1}{c}{$m_{t}$}  & \multicolumn{1}{c}{$R$} \\
\hline		
SiC	& $ 0.29 $ & $0.42$ & $1.448$\\
ZnO	& $ 0.21 $ & $0.24$ & $1.152$\\
GaN & $ 0.20 $ & $0.18$  & $0.900$\\
AlN & $ 0.33 $ & $0.25$ & $0.758$\\
\hline
\end{tabular}	
\caption{Longitudinal and transverse effective masses for some direct band gap semiconductors in terms of the free electron mass $m_{e}$.}\label{tab1}
\end{table}
	
\section{Conical geometry and the geometric potential}\label{sec3}
	
\begin{figure}[!b]
\centering
\includegraphics[scale=0.35]{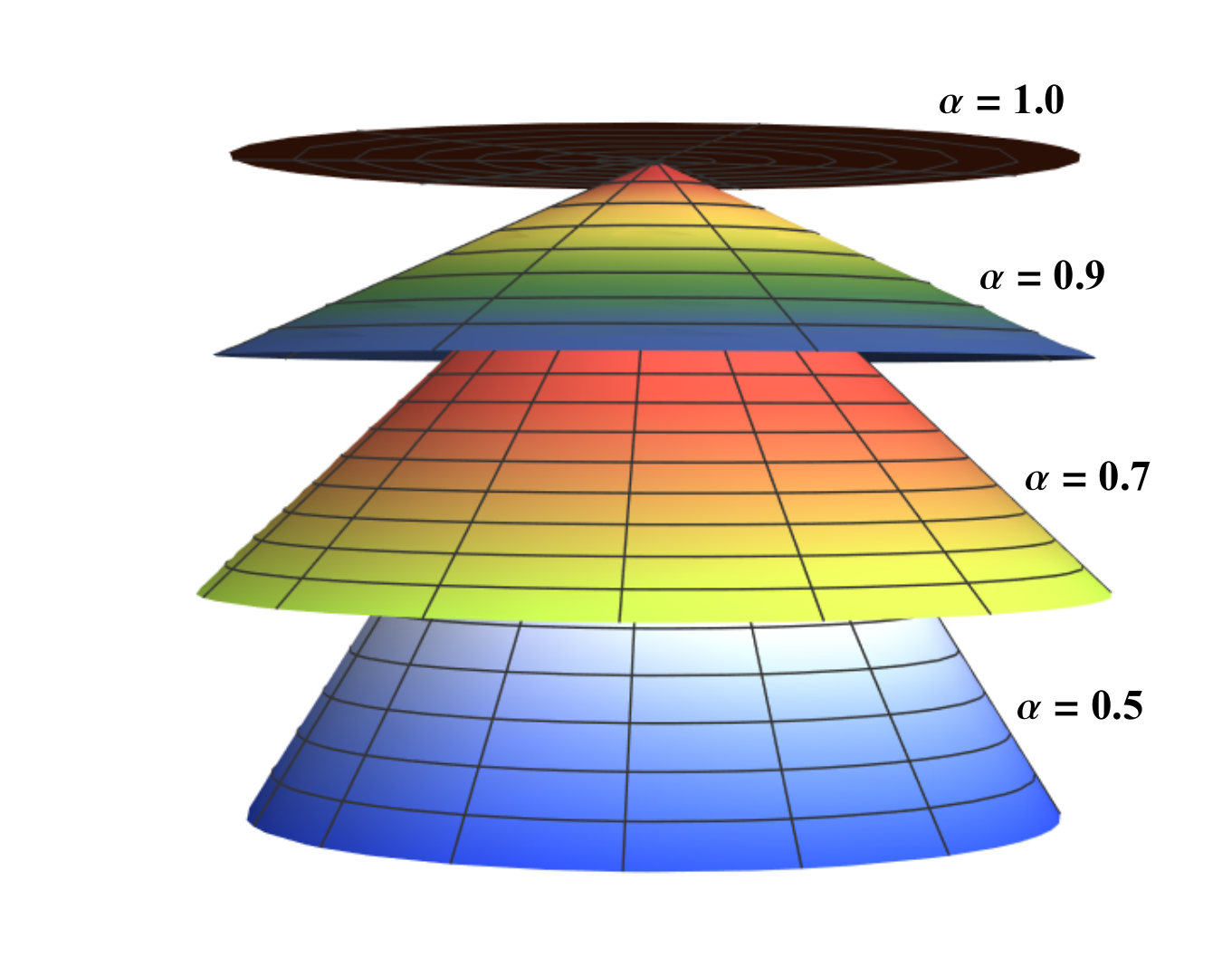}
\caption{Plot of the parametric equation (\ref{conic01}). Representation of a flat surface ($\alpha=1$) being transformed into a conical one by decreasing the $\alpha$ parameter to $0.9$, $0.7$ and $0.5$.}
\label{f1}	
\end{figure}
Since we are interested in dealing with conical defects, we begin our formalism by presenting the parametric equation of a conical surface \cite{abbena2006modern} 
\begin{equation}
\mathbf{x}(r,\theta)=(\alpha r\cos\theta, \alpha r\sin\theta, -r\sqrt{1-\alpha^2}),\label{conic01}
\end{equation}
where $r\in\mathbb{R^{+}}$ and $0\le\theta\le 2\pi$ are the surface parameters representing the distance between a given point to the origin (conic apex) and the angular coordinate, respectively. The $\alpha$ parameter designates the proportional remaining area of the original surface cut by an angular sector $\beta< 2\pi$ through the relation
	\begin{equation}
		\alpha=1-\dfrac{\beta}{2\pi}
		\label{conic02}.
	\end{equation}
Thus, when we have $\alpha=1$, Eq. (\ref{conic01}) describes a complete circle. However, for $\alpha<1$, the surface gets an angular deficit $\beta/2\pi$ leading to a conical shape as described in Fig. \ref{f1}. 
	
The geometry of a surface is determined by the coefficients of the first fundamental form \cite{differentialgeometry}, $ E=\mathbf{x}_{r}\cdot\mathbf{x}_{r}=1$, $F=\mathbf{x}_{r}\cdot\mathbf{x}_{\varphi}=0$ and $ G=\mathbf{x}_{\varphi}\cdot\mathbf{x}_{\varphi}=\alpha^{2}r^{2}$, which lead to the line element
	\begin{equation}
		ds^{2}=g_{ij}x^{i}x^{j}=dr^{2}+\alpha^{2}r^{2}d\varphi^{2},
		\label{lineelement}
	\end{equation}
with $g_{ij}$ being its metric tensor; and by those of the second fundamental form, $e=\mathbf{x}_{rr}\cdot\mathbf{N}=0$, $f=\mathbf{x}_{r\varphi}\cdot\mathbf{N}=0$ and $g=\mathbf{x}_{\varphi\varphi}\cdot\mathbf{N}=-\alpha r\sqrt{1-\alpha^{2}}$, with $\mathbf{N}=(\sqrt{1-\alpha^{2}}\cos\varphi, \sqrt{1-\alpha^{2}}\sin\varphi, \alpha)$ denoting the unitary normal vector. From the da Costa procedure \cite{PRA.1981.23.1982} of constraining a quantum particle to a surface, a purely geometric potential $V_{S}$ arises in the two-dimensional Schrödinger equation as a consequence of the curvature in the system. Such potential is known to be 
	\begin{equation}
		V_{S}=-\dfrac{\hbar^{2}}{2m}\left(M^{2}-K\right),
		\label{conic04}
	\end{equation}
where $M$ and $K$ are the mean and Gaussian curvatures, respectively. By determining the coefficients of the Gauss-Weingarten equations and using the Gauss-Bonnet Theorem \cite{Book.Manfredo.1976}, the geometric potential for the cone is given by the expression \cite{JMP.2012.53.122106,AoP.2015.362.739}\\
	\begin{equation}
		V_{S}=-\dfrac{\hbar^{2}}{2m}\left[\dfrac{(1-\alpha^{2})}{4\alpha^{2}r^{2}}-\left(\dfrac{1-\alpha}{\alpha}\right)\dfrac{\delta(r)}{r}\right],
		\label{conic05}
	\end{equation}
which presents an attractive contribution that falls with $1/r^2$ directed to the conic apex, where the potential exhibits a singular behavior represented by the Dirac delta function. 
	
	\section{Schrödinger equation: energy states and subbands}\label{sec4}
	
\begin{figure}[!h]
\centering
\includegraphics[scale=0.75]{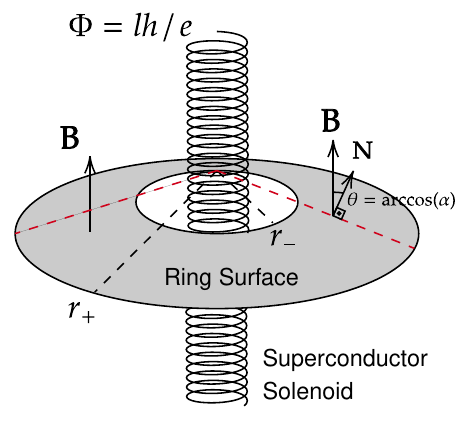}
\caption{Conical ring-shaped with a superconducting solenoid passing through its center and immersed in a uniform magnetic field $\mathbf{B}=B\hat{z}$. Because of the conical shape of the surface, the magnetic field $B$ is no longer parallel to the vector $\mathbf{N}$. $r_{-}$ and $r_{+}$ are the internal and external radius of the ring, respectively, and are measured from the idealized apex of the cone.}
\label{conical2}	
\end{figure}

In this section, we obtain the Schrödinger equation in the geometry described by the line element (\ref{lineelement}). Before proceeding, let us first introduce the model. We consider a conical ring-shaped surface (where the electrons are confined) with an AB flux tube passing through its center and immersed in a uniform magnetic field $B$ parallel to the flux tube (Fig. \ref{conical2}). The superconducting solenoid carries a quantized magnetic flux $\Phi=lh/e$.
	
The potential responsible for confining the particles in a ring-shaped region is given by \cite{SST.1996.11.1635,PRB.1999.60.5626}
	\begin{equation}
		V(r)=\dfrac{a_{1}}{r^{2}}+a_{2}r^{2}-V_{0},
		\label{sch01}
	\end{equation}
where $V_{0}=2\sqrt{a_{1}a_{2}}$. This radial potential has a minimum at $r_{0}=(a_{1}/a_{2})^{1/4}$. The quantum ring's radius and width can be adjusted independently by suitably choosing $a_{1}$ and $a_{2}$ parameters. If $r \rightarrow r_{0}$, we recover the parabolic potential,
	\begin{equation}
		V(r)=\frac{1}{2}\mu \omega_{0}\left(r-r_{0}\right)^{2}, 
		\label{Eq:potencialparabolico}
	\end{equation}
where $\omega_{0}=\sqrt{8a_{2}/\mu}$ characterizes the strength of the transverse confinement. Furthermore, in particular limits,
the radial potential can describe a quantum dot, a quantum anti-dot, a two-dimensional wire, or even a 1D ring.
	
The effective vector potential is a superposition of two vector potentials, which are responsible for the appearance of the uniform magnetic field and the AB potential, respectively. This potential is written as \cite{EPJP.2014.129.100}
	\begin{equation}
		\mathbf{A}=\dfrac{B}{2}\hat{e}_{\varphi}+\dfrac{\Phi}{2\pi\alpha^{2}r^{2}}\hat{e}_{\varphi},
		\label{effectivepotential}
	\end{equation}
where $\hat{e}_{\varphi}=\alpha r\hat{\varphi}$ is the axial base vector in conical geometry. Note that the result for the curl applied in $\alpha (Br/2) \hat{\varphi}$ is equal to $B\alpha \mathbf{N}$ which is just the component of $B$ in the normal direction since $\alpha=\cos\theta$. On Eq. (\ref{effectivepotential}), $B$ is the external uniform magnetic field, and $\Phi=\ell \Phi_{0}$ is the magnetic flux piercing through the center of the ring, where $\Phi_{0}=h/e$ is the quantum flux.
	\begin{figure*}[!t]
		\centering
		\begin{subfigure}{0.4\textwidth}
			\includegraphics[width=\textwidth]{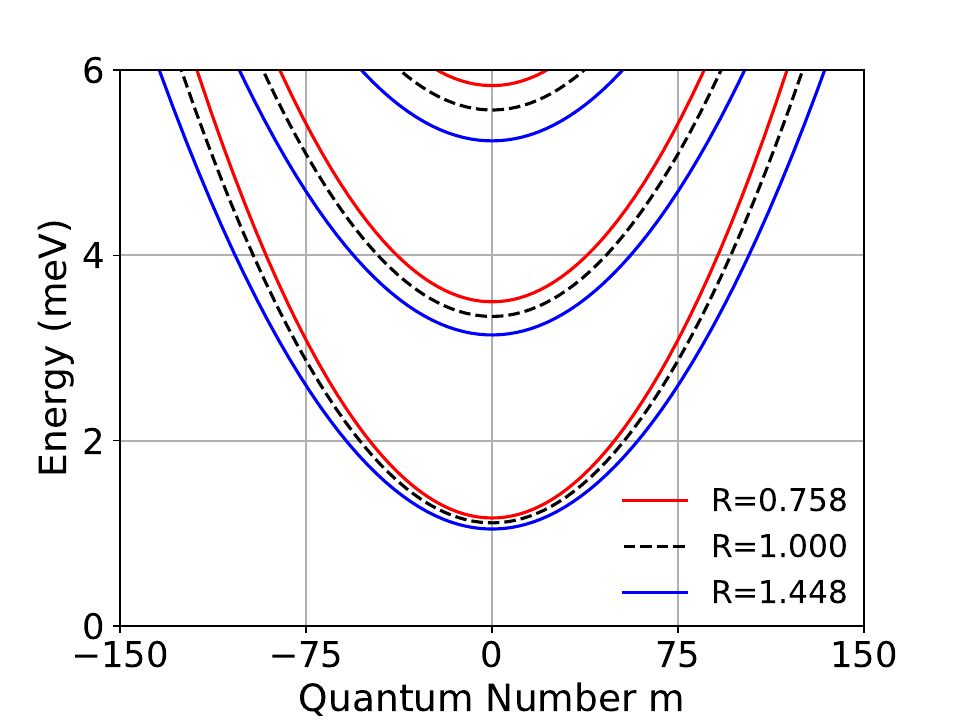}
			\subcaption{$B=0\hspace{0.05cm}T $ and $\alpha=1$}
			\label{energyma}
		\end{subfigure}
		\begin{subfigure}{0.4\textwidth}
			\includegraphics[width=\textwidth]{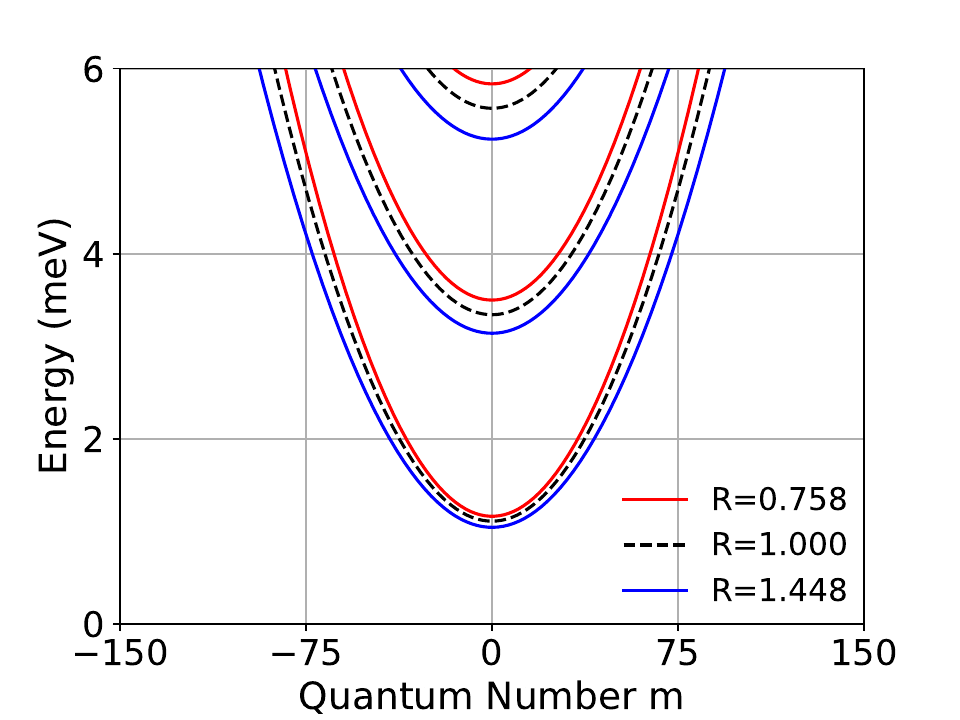}
			\subcaption{$ B=0\hspace{0.05cm}T $ and $\alpha=0.7$}
			\label{energymb}
		\end{subfigure}
		\begin{subfigure}{0.4\textwidth}
			\includegraphics[width=\textwidth]{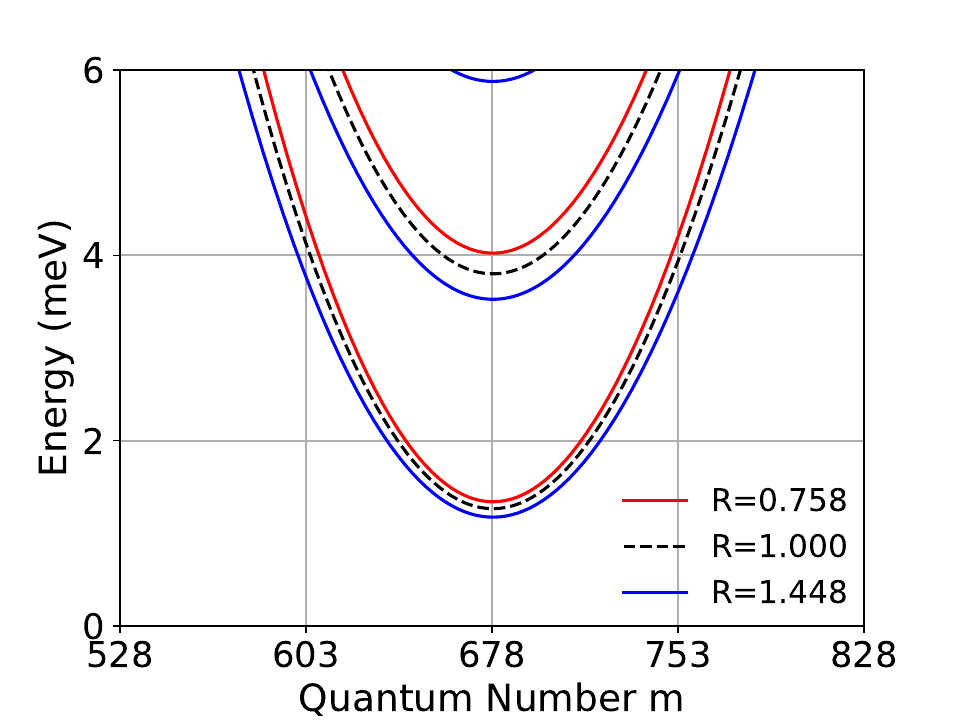}
			\subcaption{$ B=1.0\hspace{0.05cm}T $ and $\alpha=0.7$}
			\label{energymc}
		\end{subfigure}
		\begin{subfigure}{0.4\textwidth}
			\includegraphics[width=\textwidth]{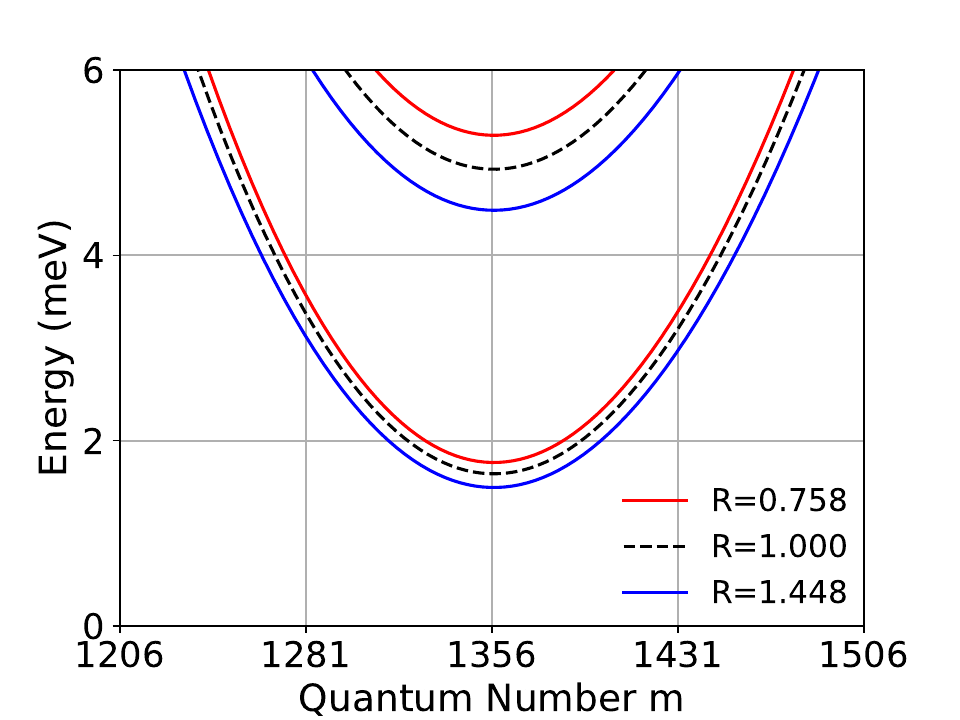}
			\subcaption{$ B=2.0\hspace{0.05cm}T $ and $\alpha=0.7$}
			\label{energymd}
		\end{subfigure}
		\caption{Subbands of the system as a function of the $m$ quantum number for different magnetic field strength and curvature values cases. According to Table 1, we apply the anisotropic ratios $R$ corresponding to the SiC and AlN semiconductors.}
		\label{energymconic}
	\end{figure*}
	\begin{figure*}[!t]
		\centering
		\begin{subfigure}{0.4\textwidth}
			\includegraphics[width=\textwidth]{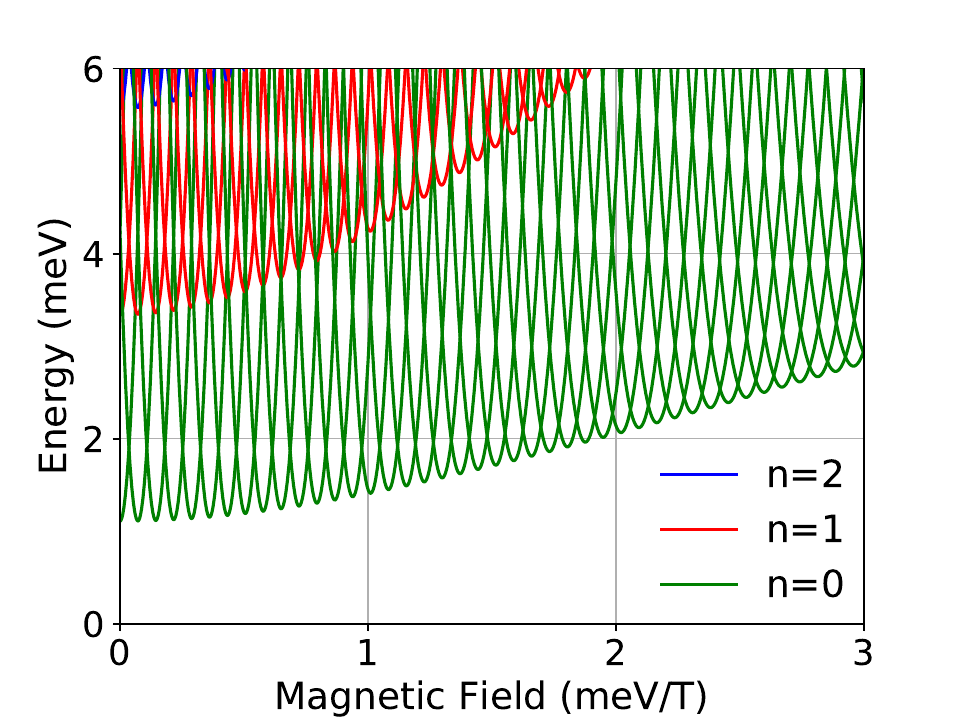}
			\subcaption{$\alpha=1$ and $R=1$}
			\label{mina}
		\end{subfigure}
		\begin{subfigure}{0.4\textwidth}
			\includegraphics[width=\textwidth]{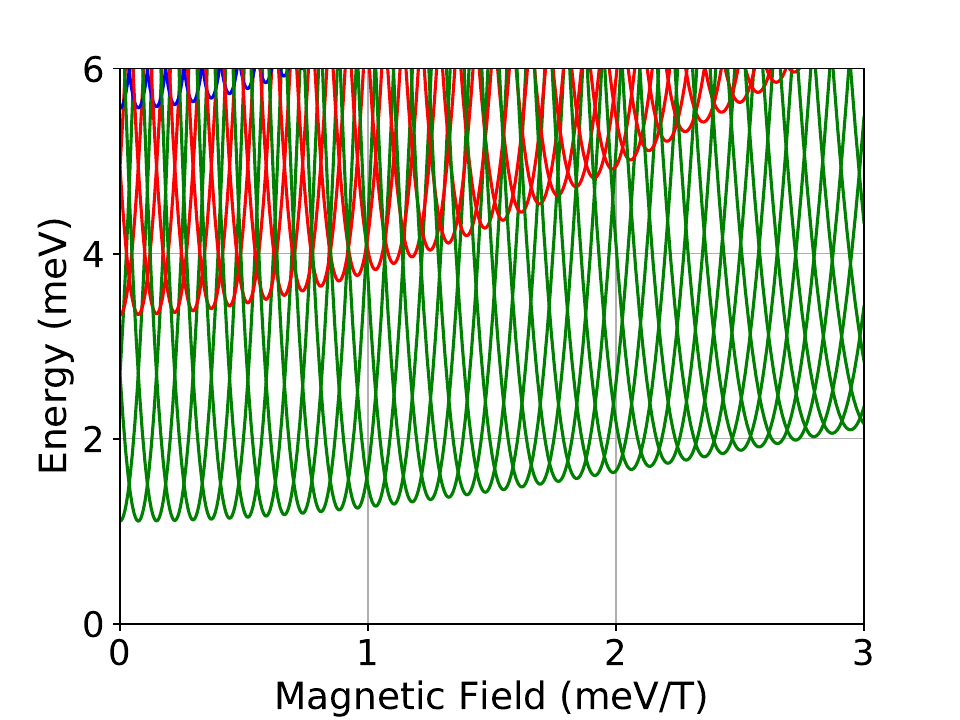}
			\subcaption{$\alpha=0.7$ and $R=1$}
			\label{minb}
		\end{subfigure}
		\begin{subfigure}{0.4\textwidth}
			\includegraphics[width=\textwidth]{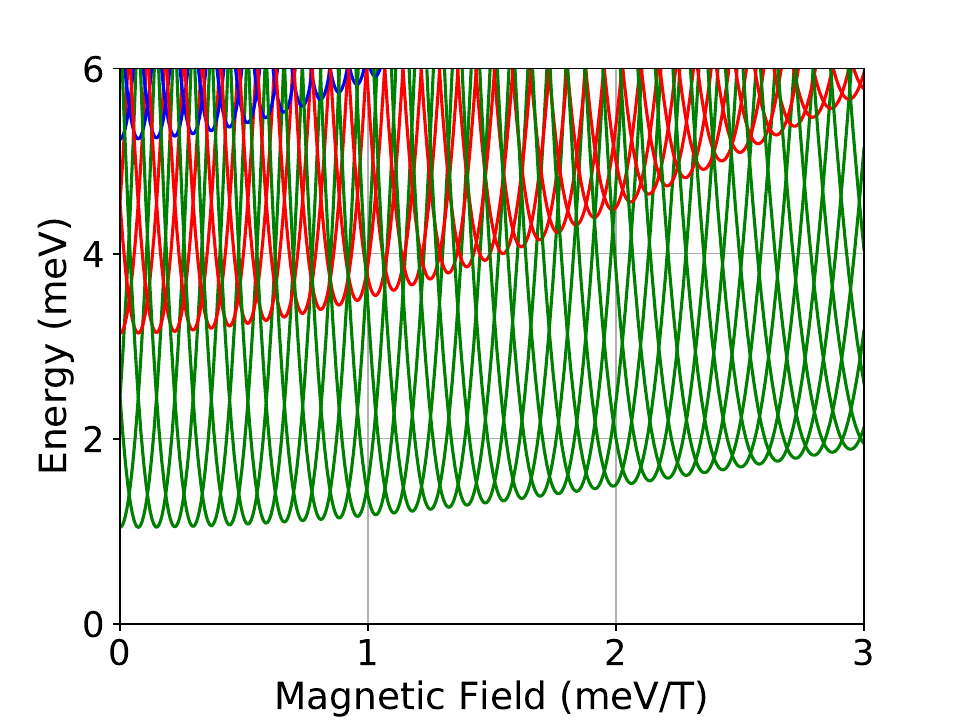}
			\subcaption{$\alpha=0.7$ and $R=1.448$}
			\label{minc}
		\end{subfigure}
		\begin{subfigure}{0.4\textwidth}
			\includegraphics[width=\textwidth]{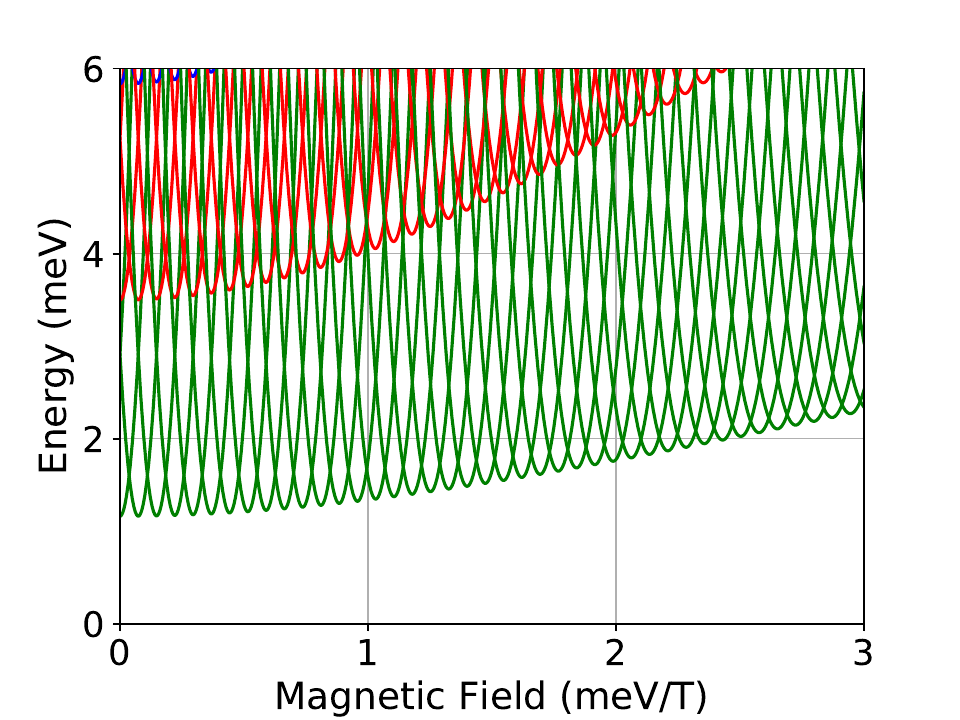}
			\subcaption{$\alpha=0.7$ and $R=0.758$}
			\label{mind}
		\end{subfigure}
		\caption{Energy of the states as a function of $B$. Frontiers between two color regions outline a subband minimum: for instance, the subband minimum for $n=1$ is the frontier between green and red regions.}
		\label{minimumsubband}
	\end{figure*}
The Schrödinger equation describing the above system has already been derived in Eq. (13) of Ref. \cite{PRL.2008.100.230403}, after including the ring potential (\ref{sch01}) to obtain
	\begin{align}
	&-\dfrac{\hbar^{2}}{2m_{t}}\left[\dfrac{1}{\sqrt{g}}\dfrac{\partial}{\partial q^{i}}\left(\sqrt{g}g^{ij}\dfrac{\partial\chi_{S}}{\partial q^{j}}\right)+\dfrac{ie}{\hbar\sqrt{g}}\dfrac{\partial }{\partial q^{i}}\left(\sqrt{g}g^{ij} A_{j}\right)\chi_{S}\right]\notag\\
	&-\dfrac{\hbar^{2}}{2m_{t}}\left[\dfrac{2ie}{\hbar}g^{ij}A_{i}\dfrac{\partial\chi_{S}}{\partial q^{j}}-\dfrac{e^{2}}{\hbar^{2}}g^{ij}A_{i}A_{j}\chi_{S}\right]\notag\\
	&-V_{S}\chi_{S} +\left(\dfrac{a_{1}}{r^{2}}+a_{2}r^{2}-V_{0}\right)\chi_{S}=	E\chi_{S},
	\label{sch03}
	\end{align}
with $\chi_{S}=\chi_{S}(r,\varphi)$ being the surface wavefunction resulting from the da Costa procedure and $i,j=\{1,2\}$. Substituting Eqs. (\ref{conic05}) and (\ref{effectivepotential}) into the Eq. (\ref{sch03}), we obtain
	\begin{align}
	&-\dfrac{\hbar^{2}}{2m_{t}}\left[\frac{1}{r}\frac{\partial}{\partial r}\left(r\dfrac{\partial}{\partial r}\right)+\frac{1}{\alpha^{2}r^{2}}\left(\frac{\partial}{\partial \varphi}+il\right)^{2}\right]\chi_{S}\notag \\
	&-\dfrac{\hbar^{2}}{2m_{t}}\left[\dfrac{ieB}{\hbar}\left(\frac{\partial}{\partial \varphi}+il\right)+\dfrac{e^{2}B^{2}\alpha^{2}r^{2}}{4\hbar^{2}}\right]\chi_{S}\notag \\ &-\dfrac{\hbar^{2}}{2m_{l}}\left[\frac{(1-\alpha^{2})}{4\alpha^{2}r^{2}}-\left(\dfrac{1-\alpha}{\alpha}\right)\frac{\delta(r)}{r}\right]\chi_{S}\notag \\
	&+\left(\dfrac{a_{1}}{r^{2}}+a_{2}r^{2}-V_{0}\right)\chi_{S}=E\chi_{S},\label{sch04}
	\end{align}
where the longitudinal mass $m_{l}$ is present in the geometric potential factor since it was only reminiscent of the normal coordinate part. Using solutions of the form $\chi_{S}(r,\varphi)=e^{-im\varphi}f(r)$ in the previous expression, we get  the radial differential equation
	\begin{align}
		&\dfrac{d^{2}f}{dr^{2}}+\dfrac{1}{r}\dfrac{df}{dr}-\left[\dfrac{2m_{t}a_{2}}{\hbar^{2}}+\dfrac{e^{2}B^{2}\alpha^{2}}{4\hbar^{2}}\right]r^{2}f(r)\notag\\
		&-\dfrac{1}{r^{2}}\left[\dfrac{2m_{t}a_{1}}{\hbar^{2}}+\dfrac{(m-l)^{2}}{\alpha^{2}}-\dfrac{(1-\alpha^{2})}{4\alpha^{2}}\dfrac{m_{t}}{m_{l}}\right]f(r)\notag\\
		&-\left[\dfrac{eB(m-l)}{\hbar}+\dfrac{2m_{t}(E+V_{0})}{\hbar^2}\right]f(r)\notag \\
		&+\dfrac{m_{t}}{m_{l}}\dfrac{(1-\alpha)}{\alpha r}\delta(r)f(r)=0,
		\label{sch04}
	\end{align}
from which we identify 
	\begin{equation}
		\dfrac{2m_{t}a_{2}}{\hbar^{2}}+\dfrac{e^{2}B^{2}\alpha^{2}}{4\hbar^{2}}=\dfrac{1}{4\lambda^{4}}\quad\text{and}\quad\omega=\sqrt{\omega^{2}_{0}+(\alpha\omega_{c})^{2}},\label{sch05}
	\end{equation}
where $\omega_{0}=\sqrt{8a_{2}/m_{t}}$, $\omega_{c}=eB/m_{t}$ is the cyclotron frequency and $\lambda=\sqrt{\hbar/m_{t}\omega}$ is the effective magnetic length. Now, by defining the following suitable quantities:
	\begin{equation} 
		L=\sqrt{\dfrac{2m_{t}a_{1}}{\hbar^{2}}+\dfrac{(m-l)^{2}}{\alpha^{2}}-\dfrac{(1-\alpha^{2})}{4\alpha^{2}}\dfrac{m_{t}}{m_{l}}}\label{sch06}
	\end{equation}
and 
	\begin{equation}
		S=\sqrt{\dfrac{eB(m-l)}{\hbar}+\dfrac{2m_{t}(E+V_{0})}{\hbar^2}},\label{sch07}
	\end{equation}
we stay with a compact form for the differential equation (\ref{sch04})
	\begin{align}
		\frac{d^{2}f(r)}{dr^{2}}+\frac{1}{r}\frac{df(r)}{dr}-&\left[\frac{r^{2}}{4\lambda^{4}}+\frac{L^{2}}{r^{2}}-S^{2}\right]f(r)\notag \\ &-\frac{m_{t}}{m_{l}}\frac{(1-\alpha)}{\alpha r}\delta(r)f(r)=0.
		\label{sch08}
	\end{align}
Let us write Eq. (\ref{sch08}) in the form $hf(r)=Sf(r)$, where $h$ is the Hamiltonian operator. We emphasize that the operator $h$ is essentially self-adjoint for $\left|L\right| \geq 1$, while if $\left|L\right|< 1$, it is not self-adjoint \cite{Book.2004.Albeverio,PRD.2012.85.041701}.
The presence of the $\delta$ function in the radial equation above implies that the Hamiltonian is not a self-adjoint operator. In this approach, the singular solutions should be considered, and the energy spectrum is obtained through the self-adjoint extension method \cite{PRL.1990.64.503,PRD.2012.85.041701}. Nevertheless, for a quantum ring in the mesoscopic regime, we can infer that $|L|>1$. Thus, we can neglect the influence of the $\delta$ function and consider only regular solutions of the radial equation. With a suitable change of variables, it can be found from Eq. (\ref{sch08}) a confluent hypergeometric-type differential equation from which we can obtain the energy eigenvalues of the system. In this manner, the wavefunctions and the corresponding energy eigenvalues of the Eq. (\ref{sch08}) are given by
\begin{align}
\chi_{n,m}&(r,\varphi)=\dfrac{1}{\lambda}\sqrt{\dfrac{\Gamma(L+n+1)}{2^{L+1}\pi\{\Gamma(L+1)\}^2 n!}}\left(\frac{r}{\lambda}\right)^{L}\notag\\&\times e^{-1/4(r/\lambda)^2}\, _{1}F_{1}\left(-n, L+1, \dfrac{1}{2}(r/\lambda)^2\right)e^{-im\varphi},
\label{sch09}
\end{align}
\begin{equation}
E_{n,m}=\left(n+\dfrac{1}{2}+\dfrac{L}{2}\right)\hbar\omega-\dfrac{(m-l)}{2}\hbar\omega_{c}-\dfrac{m_{t}}{4}\omega_{0}^{2}r^{2}_{0},
\label{sch10}
\end{equation}
where $_{1}F_{1}$ is the confluent hypergeometric function of the first kind, and $n=0,1,2,3,\dots$  is the radial quantum number, which denotes the subband level. 
We can note that the $\alpha$ parameter directly modulates the influence of the magnetic field since the flux of $B$ through the surface enclosed by the ring reduces as $\alpha\rightarrow 0$. For $\alpha=1$, we recover the results from \cite{PRB.1999.60.5626} without anisotropy.

	\section{Energy levels}\label{sec5}
	
In this section, we study the effects of anisotropy, curvature, and magnetic field on the electronic properties. Our analysis considers a sample containing $1400$ spinless and non-interacting electrons in a ring-shaped region of average radius $r_{0}= 1350$ nm. The confinement energy is $\hbar\omega_{0}=2.23$ meV. 
	
Figure \ref{energymconic} shows the energy eigenvalues as a function of $m$ for the anisotropy parameter $R$ of SiC and AlN semiconductors in different magnetic fields and curvature regimes. 
	\begin{figure}[!t]
		\centering
		\includegraphics[scale=0.48]{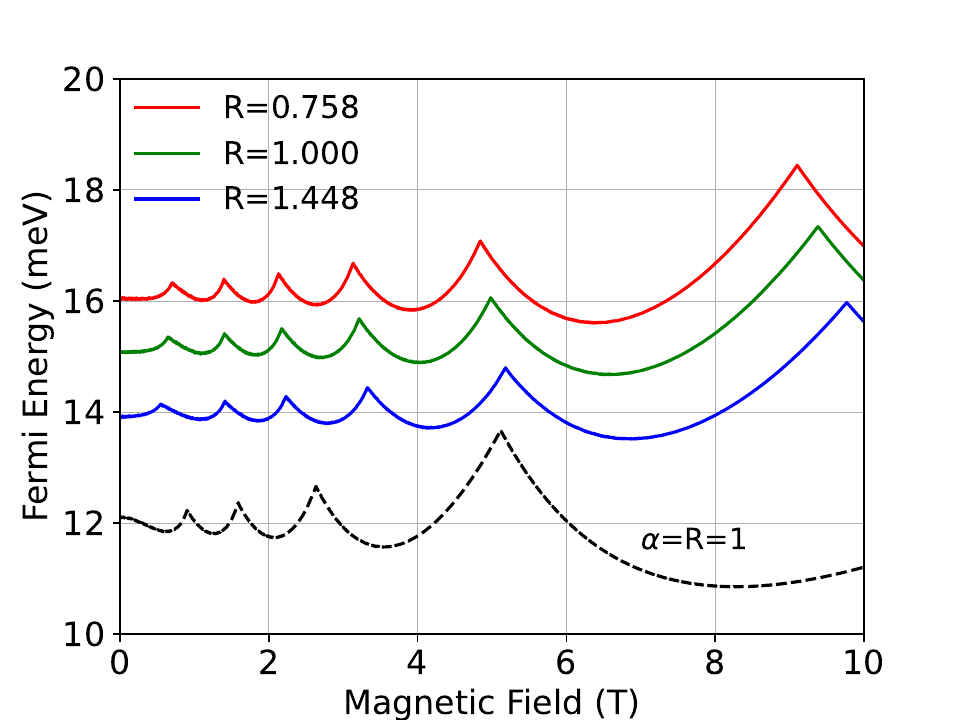}
		\caption{Fermi energy of the system as a function of the magnetic field for $\alpha=0.7$ and different values of $R$. The dashed-line curve represents the flat and isotropic case.}
		\label{fermienergy}	
	\end{figure}
	\begin{figure}[t]
		\centering
		\includegraphics[scale=0.48]{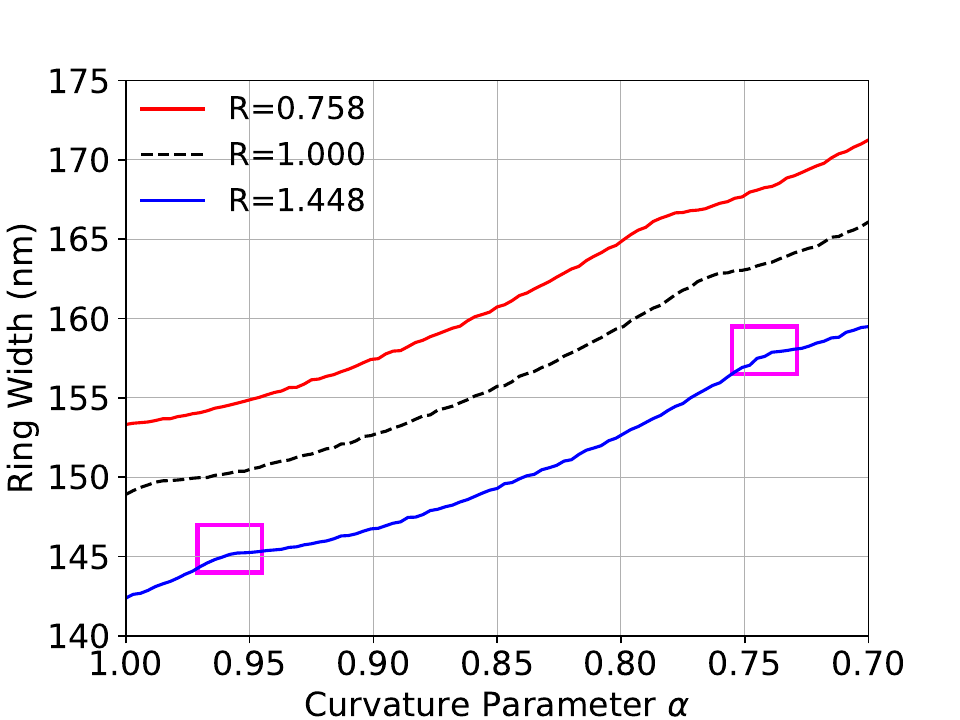}
		\caption{Quantum width of the system (difference between the external and internal radius of the ring) as a function of the curvature parameter $\alpha$ for different values of $R$. The magenta squares indicate the addition of another occupied subband to the system.}
		\label{ringwidth}	
	\end{figure}
Between Figs. \ref{energyma} to \ref{energymb}, the curvature parameter has changed to provide the conic deformation in the system, as already illustrated in Figure \ref{f1}. It gives a narrowing effect to the subbands, implying the energy lifting of the states with higher $m$ numbers. In fact, due to the presence of the term $m^2/\alpha^2$ in the effective angular momentum $L$, lower values of $\alpha$ imply a higher dispersion of the curves.
	
Anisotropy was included by assuming values of $R$ higher (blue color) or lower (red color) than $1$. In Fig. \ref{energymconic}, we see that anisotropy brings effects in the subbands by the dislocation of their bottoms upper ($R<1$) or lower ($R>1$) when compared to the isotropic case (dashed line). Figures \ref{energymc} to \ref{energymd}, the magnetic field intensifies the effect from the anisotropy, besides enlarging the subband's width.
	
As can be computed from Eq. (\ref{sch10}), energy separation between neighboring subbands is provided by $\hbar \omega$, where $\omega$ is given in Eq. (\ref{sch05}). Interestingly, the anisotropy correction for the energy separation between neighboring subbands is inversely proportional to $R$. This result can be verified by substituting Eq. (\ref{anis03}) in Eq. (\ref{sch05}).
Thus, for $R=1$ (Figs. \ref{mina} and \ref{minb}), we can observe that the separation energy between the neighboring subbands increases with the magnetic field, as already pointed out in Refs. \cite{SST.1996.11.1635}. 
One can see that this effect has its intensity reduced by including curvature \cite{AdP.2019.531.1900254}. Furthermore, anisotropy can modulate this dependence: For $R>1$ (Fig. \ref{minc}), the effect of this magnetic field-dependent separation is reduced, while the $R<1$ case (Fig. \ref{mind}) provides a growth.   
	
	\section{Fermi energy and quantum width}\label{sec6}
	
The Fermi energy $\epsilon_{f}$ of the quantum ring for a given magnetic field magnitude is the highest energy value that a state of the system with $N$ spinless and non-interacting electrons can achieve in the absolute zero. At this temperature, the electrons accommodate themselves in the lowest available energy states until the complete population. Thus, each subband's extreme quantum number value $m$ will be limited to the $\epsilon_{f}$ value. The Fermi energy can be computed self-consistently using the following expression:
	\begin{equation}
		N_{e}=\sum_{n,m}\theta \left(\epsilon_{F}-E_{n,m} 
		\right),
		\label{Eq:EnergiaFermi}    
	\end{equation}
where $N_{e}$ is the total number of electrons and $E_{n,m}$ is given by Eq. (\ref{sch10}). $\theta\left(\epsilon_{F}-E_{n,m} 
\right)$ represents de Heaviside function returning values $1$ when $\epsilon_{F}\ge E_{n,m}$ and $0$ otherwise. We use the numerical approach to find the $\epsilon_{F}$ value for each $B$.
	
The Fermi energy is an essential quantity for the thermodynamic properties of the system. It is closely related to the internal energy and the quantum pressure in other mesoscopic systems, such as quantum dots, where the variation in their sizes influences their optical properties substantially. As we shall see later, Fermi energy also determines the magnetization behavior.
	
In Fig. \ref{fermienergy}, we plot the Fermi energy curves for different values of the anisotropy parameter. For
simplicity, we investigated only the case $\alpha=0.7$. Behavior for $\alpha=1$ is also outlined for comparison \cite{PRB.1999.60.5626}. As the value of R decreases (increases), the Fermi energy curves experiment a lifting (reduction), and their respective peaks are shifted to lower (higher) values of magnetic fields due to the anisotropy effects in the subbands, as discussed in Section \ref{sec5}.
	
When the sample is curved, the radial confinement potential of the ring will be complemented by the non-zero term of geometric potential (\ref{conic05}), constituting the effective scalar potential to which the 2DEG is subjected. Then, we have
	\begin{equation}
		V(r)=\dfrac{a_{1}}{r^2}+a_{2}r^2-2\sqrt{a_{1}a_{2}}-\dfrac{\hbar^{2}}{2m_{l}}\dfrac{(1-\alpha^{2})}{4\alpha^{2}r^{2}},
		\label{fw01}
	\end{equation}
where we again impute the longitudinal effective mass for the geometric potential term. Fixing $V(r)=\epsilon_{f}$ in Eq. (\ref{fw01}) and solving for the variable $r$, the difference between the two roots $r_{+}$ and $r_{-}$ gives us the quantum width of the ring $\Delta r$ whose expression is given by
	\begin{equation}
		\Delta r=r_{0}\left[\dfrac{2\epsilon_{f}}{V_{0}}+2\left(1-\sqrt{1-\dfrac{R(1-\alpha^2)}{\alpha^2}\left(\dfrac{\hbar\omega_{0}}{4V_{0}}\right)^2}\right)\right]^{\frac{1}{2}},
		\label{fw02}
	\end{equation}
with $V_{0}=2\sqrt{a_{1}a_{2}}=(m_{t}/4)\omega^2_{0}r^{2}_{0}\approx 1992.22\hspace{0.05cm}\text{meV}$ and $r_{0}\approx 1350\hspace{0.05cm}\text{nm}$ being the average radius of the ring without curvature and anisotropy (\text{GaAs} sample). Note that $V_{0}$ is constant for any value of $R$, and we can return to the result found in Ref. \cite{PRB.1999.60.5626} when $\alpha=1$ and the effective mass is isotropic. 
\begin{figure}[!h]
\centering
\includegraphics[scale=0.48]{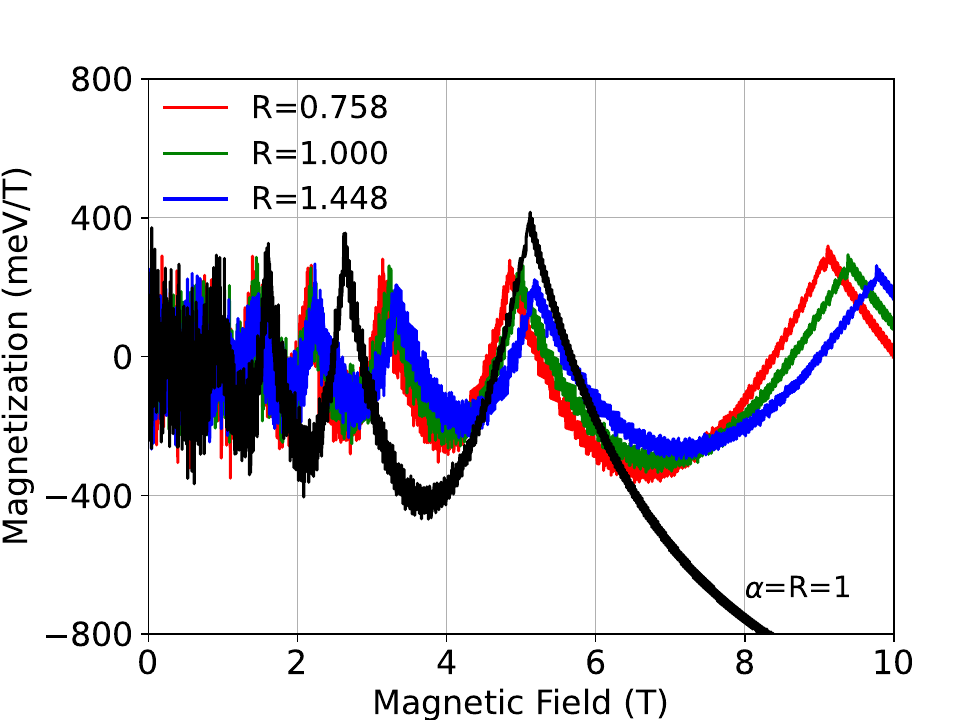}
\caption{Total magnetization of the system as a function of the magnetic field. One can see the influence of the Fermi energy behavior on the magnetization pattern such as the number of peaks and their localization in the magnetic field axis.}
\label{magnetization}	
\end{figure}
	\begin{figure*}[!t]
		\centering
		\begin{subfigure}{0.4\textwidth}
			\includegraphics[width=\textwidth]{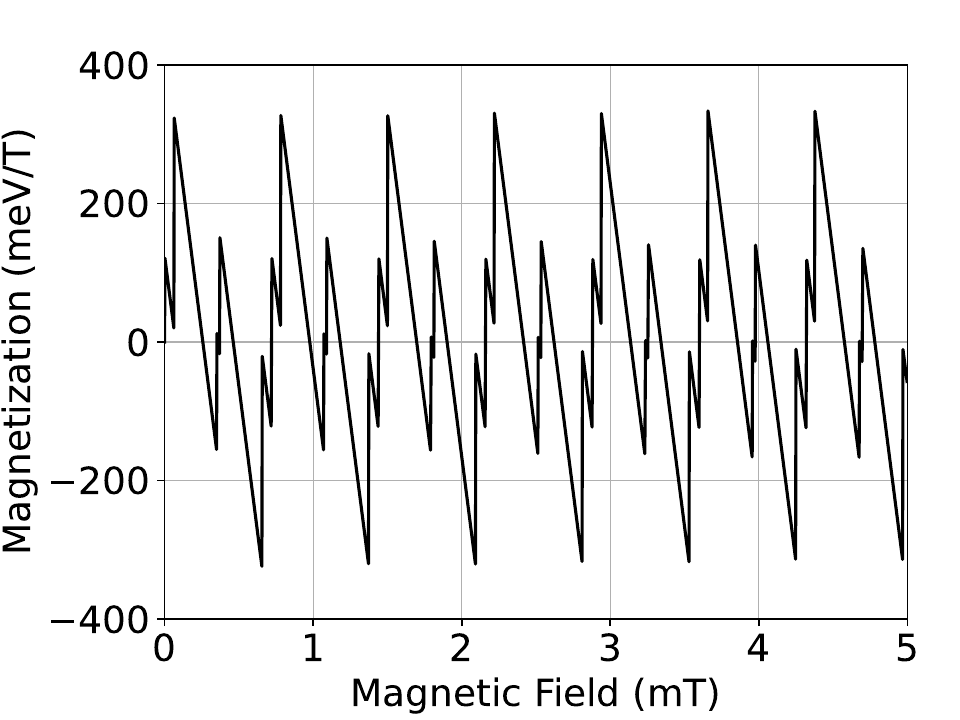}
			\subcaption{$ \alpha=R=1 $}
			\label{maglow101}
		\end{subfigure}
		\begin{subfigure}{0.4\textwidth}
			\includegraphics[width=\textwidth]{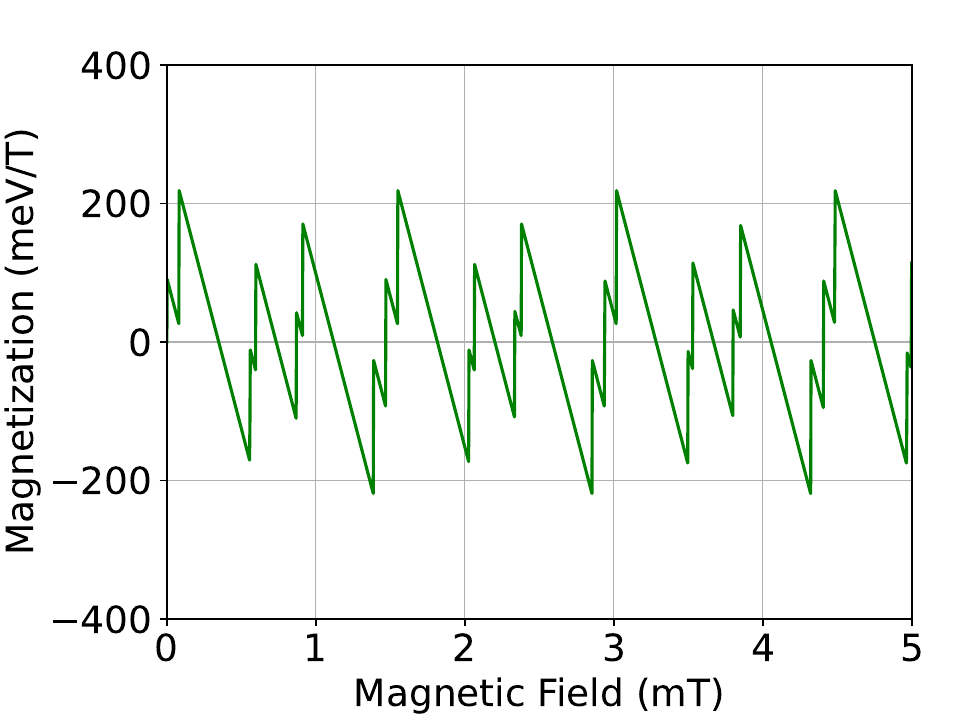}
			\subcaption{$\alpha=0.7$ and $R=1 $}
			\label{maglow107}
		\end{subfigure}
		\begin{subfigure}{0.4\textwidth}
			\includegraphics[width=\textwidth]{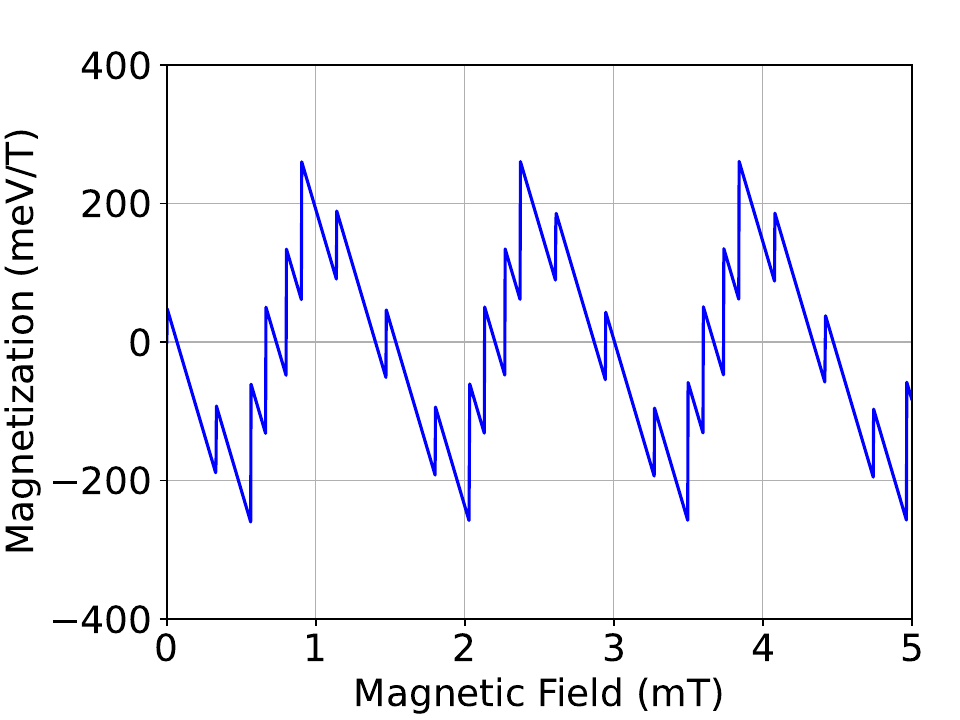}
			\subcaption{$ \alpha=0.7$ and $R=1.448$}
			\label{maglow101448}
		\end{subfigure}
		\begin{subfigure}{0.4\textwidth}
			\includegraphics[width=\textwidth]{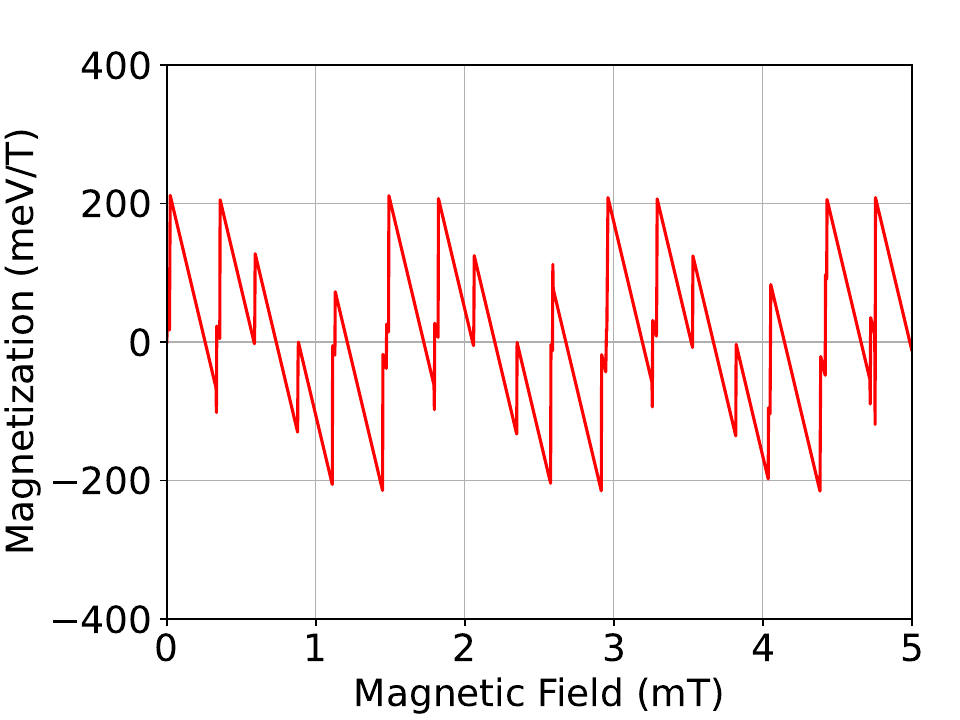}
			\subcaption{$ \alpha=0.7$ and $R=0.758 $}
			\label{maglow100758}
		\end{subfigure}
		\caption{Magnetization of the system for the low-magnetic field case. The AB oscillations in magnetization occur in periods of $0.000717\hspace{0.05cm}\text{T}/\alpha^2$. }
		\label{anismagnetizationlow}
	\end{figure*}
	\begin{figure*}[!t]
		\centering
		\begin{subfigure}{0.4\textwidth}
			\includegraphics[width=\textwidth]{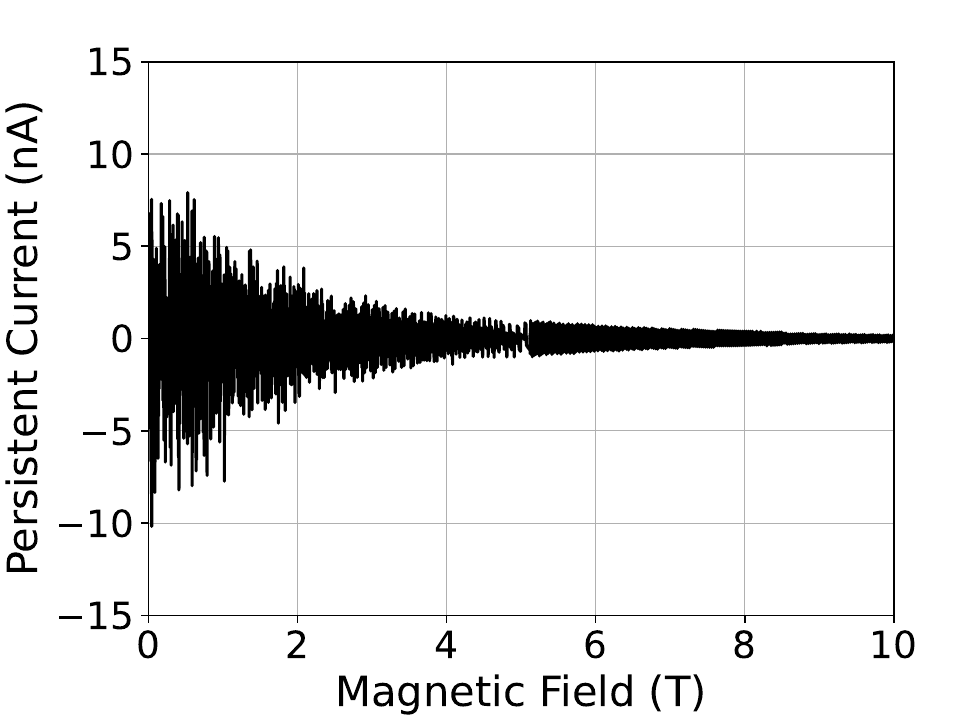}
			\subcaption{$\alpha=1$ and $R=1$}
		\end{subfigure}
		\begin{subfigure}{0.4\textwidth}
			\includegraphics[width=\textwidth]{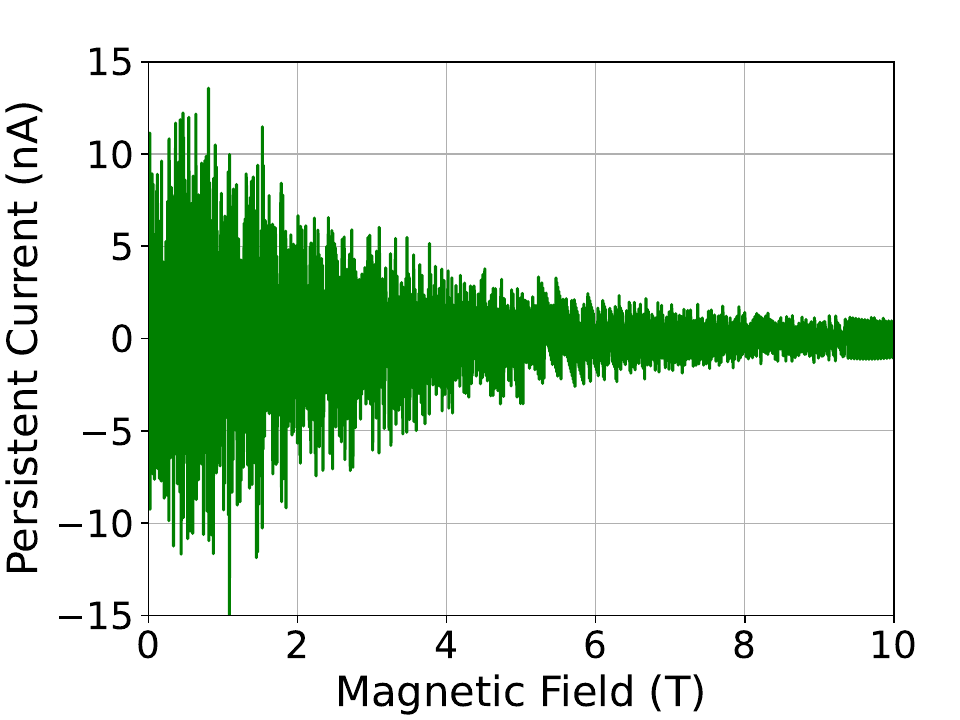}
			\subcaption{$\alpha=0.7$ and $R=1$}
		\end{subfigure}
		\begin{subfigure}{0.4\textwidth}
			\includegraphics[width=\textwidth]{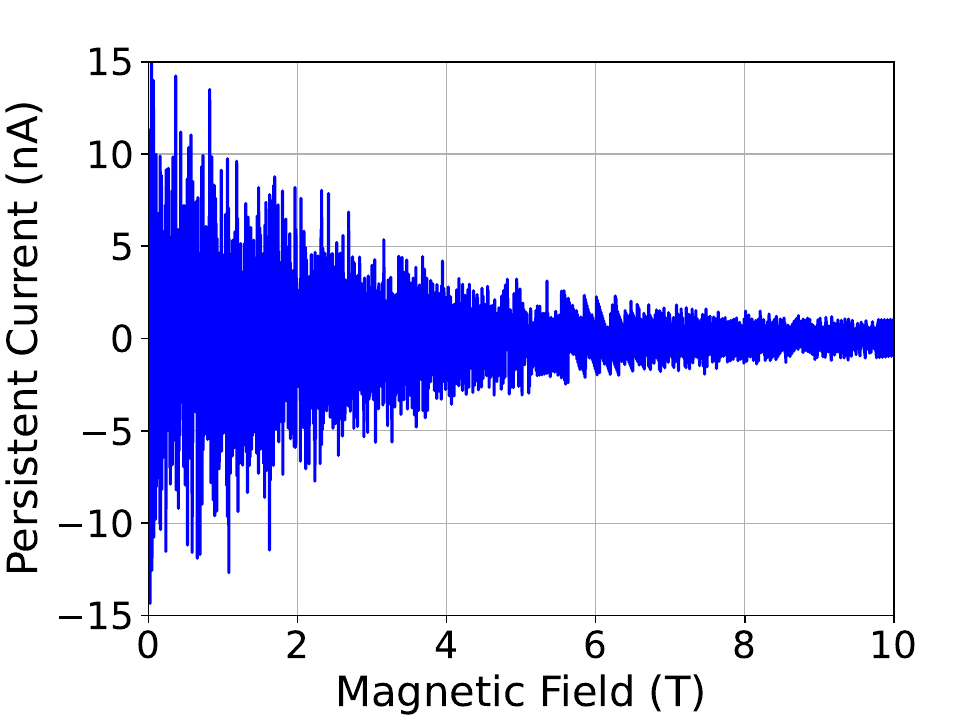}
			\subcaption{$\alpha=0.7$ and $R=1.448$}
		\end{subfigure}
		\begin{subfigure}{0.4\textwidth}
			\includegraphics[width=\textwidth]{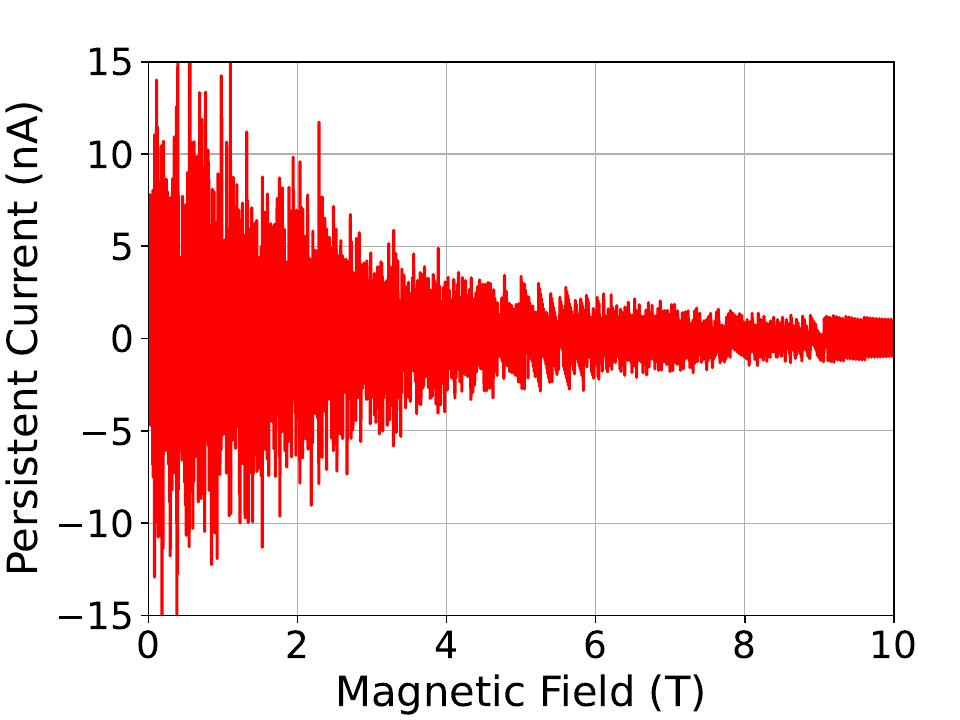}
			\subcaption{$\alpha=0.7$ and $R=0.758$}
		\end{subfigure}
		\caption{Total persistent current of the system with the magnetic field strength for different regimes of curvature and anisotropy.}
		\label{anispersi}
	\end{figure*}
	\begin{figure*}[!t]
		\centering
		\begin{subfigure}[t]{0.4\textwidth}
			\includegraphics[width=\textwidth]{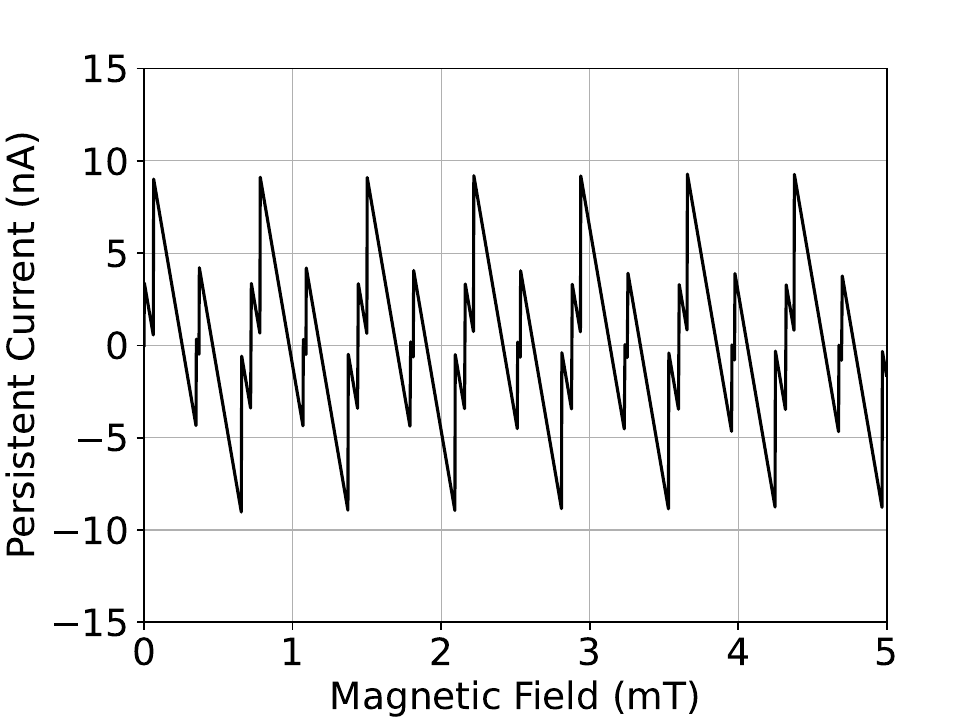}
			\subcaption{$ \alpha=1 $ and $R=1$}
		\end{subfigure}
		\begin{subfigure}[t]{0.4\textwidth}
			\includegraphics[width=\textwidth]{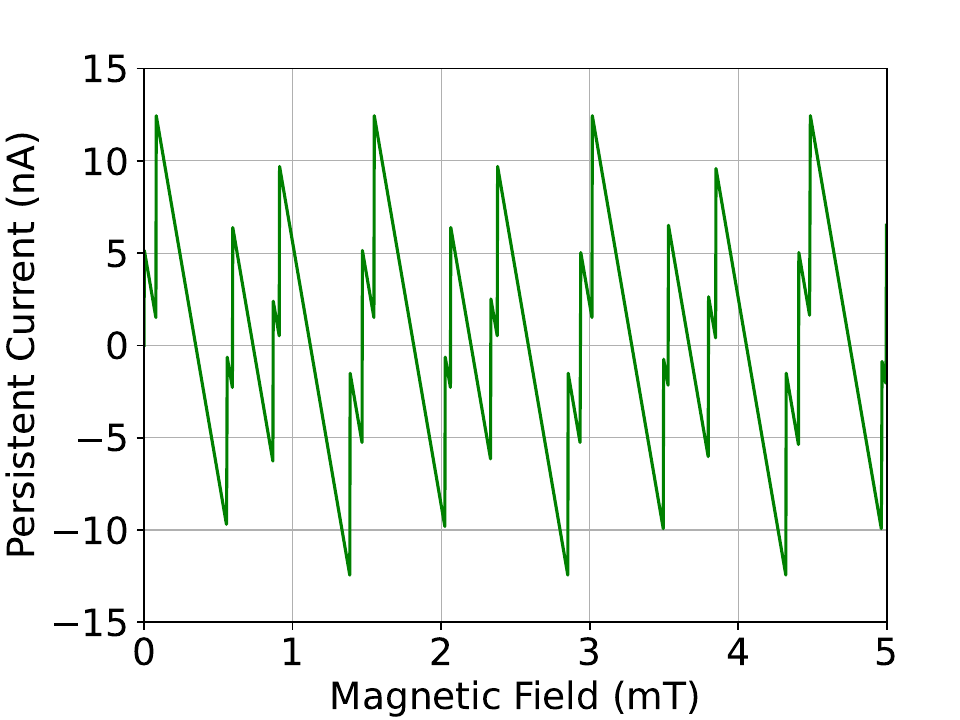}
			\subcaption{$ \alpha=0.7 $ and $R=1$}
		\end{subfigure}
		\begin{subfigure}[t]{0.4\textwidth}
			\includegraphics[width=\textwidth]{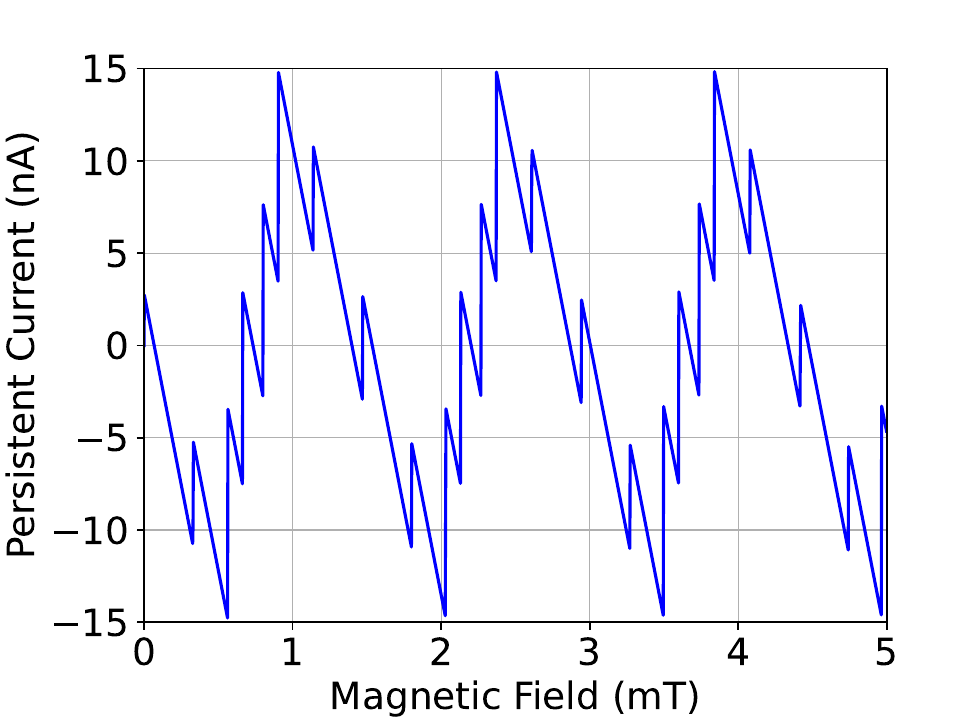}
			\subcaption{$ \alpha=0.7 $ and $R=1.448$}
		\end{subfigure}
		\begin{subfigure}[t]{0.4\textwidth}
			\includegraphics[width=\textwidth]{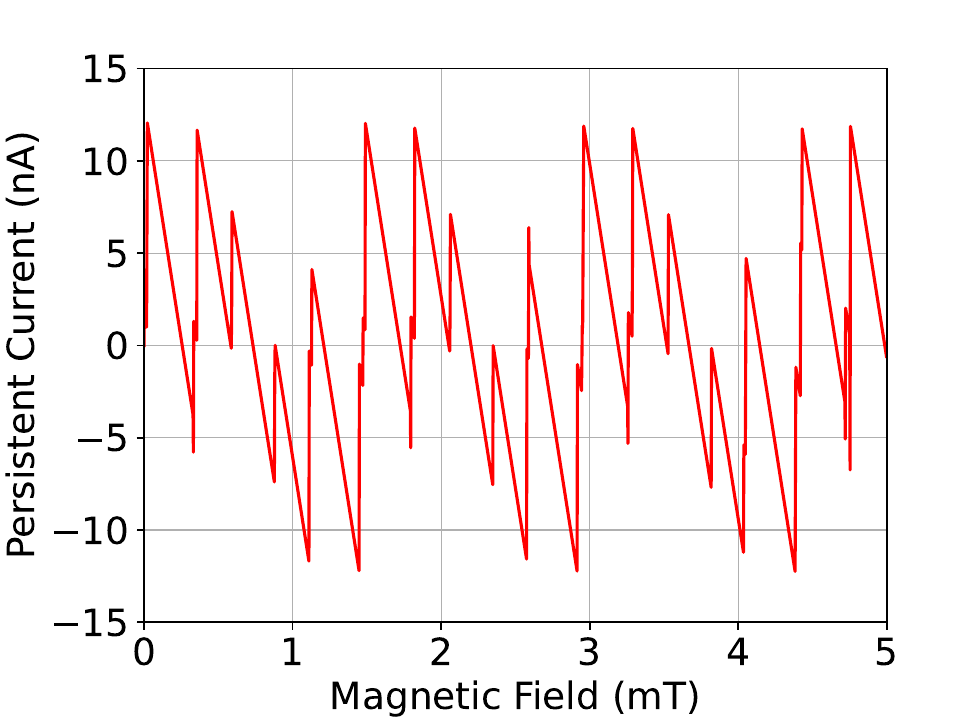}
			\subcaption{$ \alpha=1 $ and $R=0.758$}
		\end{subfigure}
		\caption{Total persistent current of the system in the weak magnetic field regime for different values of $\alpha$ and $R$ parameters. One can see both curvature and anisotropy can affect the amplitudes and period of oscillations.}
		\label{anispersilow}
	\end{figure*}
	\begin{figure}[!t]
		\centering
		\includegraphics[scale=0.48]{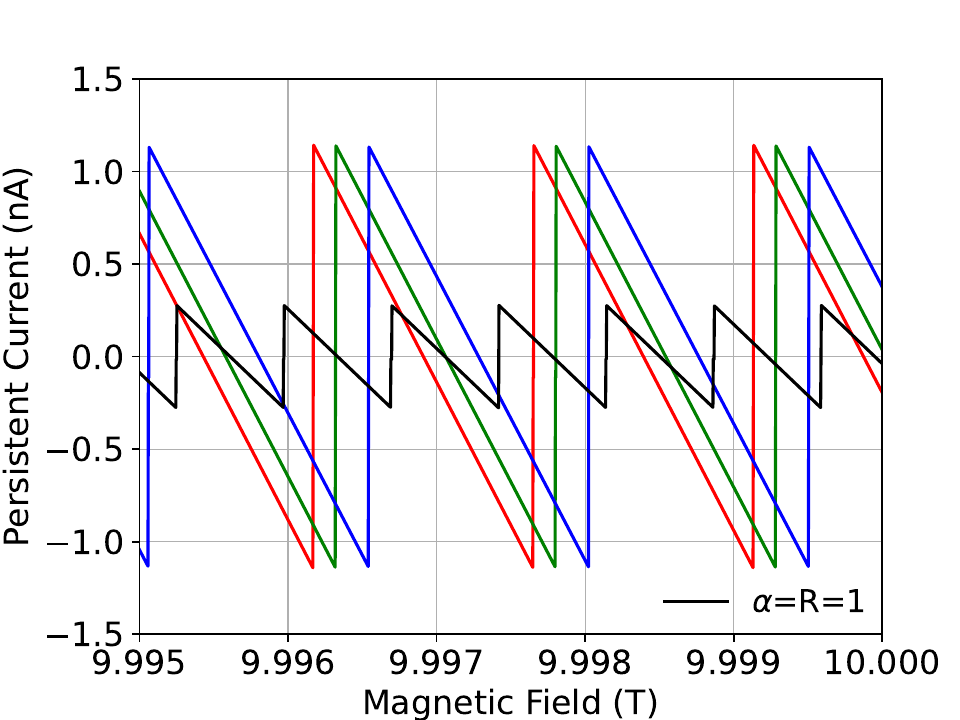}
		\caption{Total persistent current of the system in the strong magnetic field regime for different values of $\alpha$ and $R$ parameters. As expected from expression \ref{persi03}, the amplitude of the persistent currents is strongly reduced.}
		\label{persihigh}	
	\end{figure}
In Fig. \ref{ringwidth}, we plot the profile of the quantum width responding to a continuous curvature variation for different values of $R$. For the sample parameters used in this work, it is verified that $\hbar\omega_{0} << V_{0}$. As a consequence, the main contribution in expression (\ref{fw02}) must come from the Fermi energy. This result suggests one could regulate the size of the quantum rings and dots by using materials with different $R$ and $\alpha$ parameters for the same number of carriers in the 2DEG. Moreover, one could identify the existence of specific values of $\alpha$ (demarcated by the magenta squares for the $R=1.448$ case) where the derivative of the ring width curve suffers a sudden reduction. These points indicate the addition of another occupied subband and, consequently, another peak in the Fermi energy. Besides that, we see further that the increase of the prolate anisotropy ($R<1$) induces the rising of peaks for lower curvatures.     
	\begin{figure*}[!t]
		\centering
		\begin{subfigure}{0.4\textwidth}
			\includegraphics[width=\textwidth]{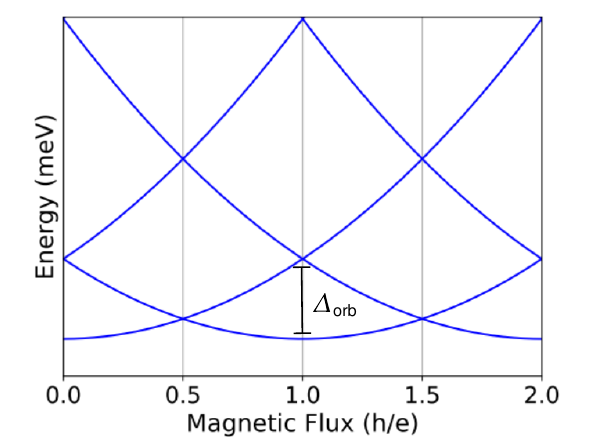}
			\subcaption{}
			\label{fluxqubit}
		\end{subfigure}
		\begin{subfigure}{0.4\textwidth}
			\includegraphics[width=\textwidth]{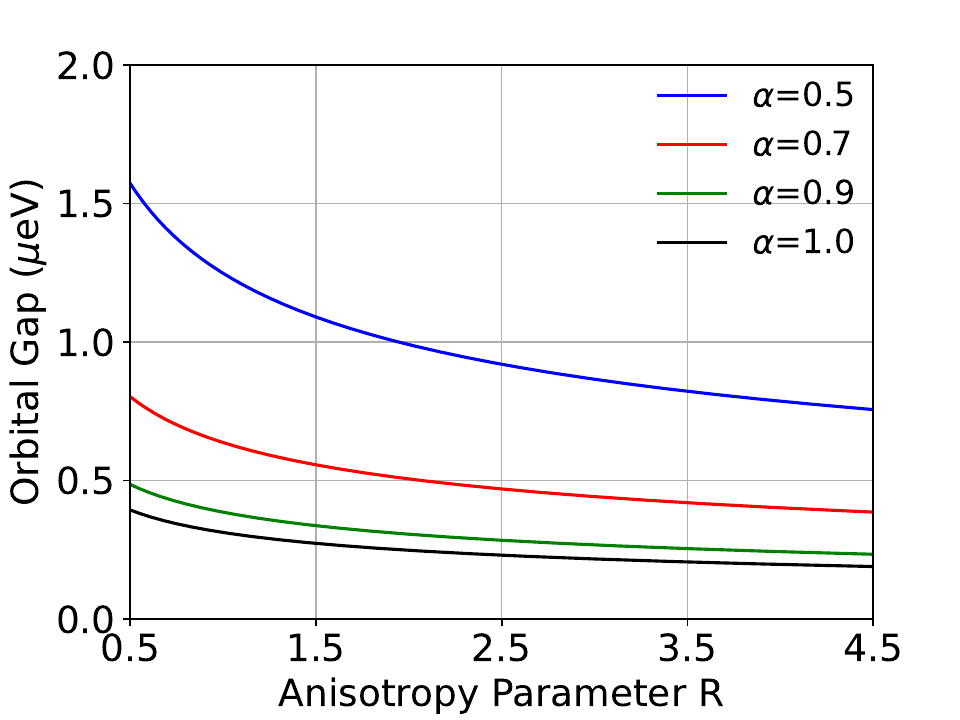}
			\subcaption{}
			\label{gaporbital}
		\end{subfigure}
		\caption{(a) Energy of the orbitals in the zero external magnetic field case when there is only magnetic flux through the solenoid. (a) The behavior of the orbital gap as a function of the anisotropic parameter for different curvature cases.}
		\label{...}
	\end{figure*}

	\section{Magnetization}\label{sec7}
	
In a thermodynamic system under a magnetic field $B$, the total magnetization $\mathcal{M}$ is an intensive parameter that determines the response of the total free energy $F$ due to the field variation \cite{thermodynamics}, and it is defined by
	\begin{equation}
	\mathcal{M}(B)=-\dfrac{\partial F}{\partial B}=-\sum^{N_{e}}_{n,m}\dfrac{\partial E_{n,m}}{\partial B}=\sum^{N_{e}}_{n,m}\mathcal{M}_{n,m},
	\label{mag01}
	\end{equation}
	with $E_{n,m}\le \epsilon_{f}$, where the state magnetization $\mathcal{M}_{n,m}$ is obtained explicitly from Eq. (\ref{sch10}) as
	\begin{equation}
	\mathcal{M}_{n,m}=-\dfrac{e\hbar}{\mu R^{1/3}}\left[\alpha^{2}\left(n+\dfrac{1}{2}+\dfrac{L}{2}\right)\dfrac{\omega_{c}}{\omega}-\dfrac{m-l}{2}\right].
	\label{mag02}
	\end{equation}
The expression (\ref{mag02}) is in good agreement with the results presented in \cite{PRB.1999.60.5626} when it is applied to the $R=\alpha=1$ case as well as the behavior of the total magnetization sketched in Figure \ref{magnetization}. The influence of the Fermi energy in the total magnetization can be seen by the occurrence number of peaks and their respective dislocations through the magnetic field axis \cite{PRB.1999.60.5626, PEREIRA2021114760}. These magnetization oscillations, characterized by the peaks, arise due to the subband depopulation and are known as dHvA oscillations. 
	
One can observe in Fig. \ref{magnetization} that the ring curvature strongly reduces their amplitudes. Moreover, the anisotropy also affects these amplitudes by decreasing them with the more anisotropic prolate ($R<1$) the system becomes. In the weak magnetic field regime, the oscillations presented in magnetization are due to the intersection of the energy states, and they are known as AB oscillations, causing the electrons to oscillate in their energies through these state transitions. AB oscillations are more sensible with the variation of $B$ when several occupied subbands exist. Since curvature increases the number of occupied subbands in the system at zero magnetic fields, more transitions per period will be shown for the cases of Figs. \ref{maglow107}, \ref{maglow101448} and \ref{maglow100758} when $\alpha=0.7$. Furthermore, the period of these oscillations $p_{0}\approx 0.000717\hspace{0.05cm}\text{T}$ observed in \cite{PRB.1999.60.5626} suffers an increase of its value to $p_{0}/\alpha^2\approx 0.001463\hspace{0.05cm}\text{T}$ when we are considering $\alpha=0.7$.   
	
	\section{Persistent currents}\label{sec8}
	
Persistent currents designate the electric charge movements through a material with a very low dissipation, a well-known phenomenon in superconductors first observed in 1911 with  Kamerlingh Onnes \cite{Kittel2004}. During the 1980s and the beginning of the 1990s, it was predicted and experimentally verified for the first time the existence of this kind of currents in non-superconductor metals \cite{BUTTIKER1983365, PhysRevLett.54.2049, PhysRevLett.64.2074, PhysRevLett.67.3578}. Moreover, in 1993, Maily et al. performed the first experiment using mesoscopic rings made of a semiconductor heterojunction of GaAs/GaAlAs \cite{PhysRevLett.70.2020} where persistent currents were observed. 
The expression for the persistent current correspondent to a specific state $I_{n,m}$ in a quantum ring is provided by the Byers-Yang relation \cite{PhysRevLett.7.46}
	\begin{equation}
	I_{n,m}=-\dfrac{1}{\phi_{0}}\dfrac{\partial E_{n,m}}{\partial l}=\dfrac{e\omega}{4\pi}\left(\dfrac{m-l}{\alpha^2 L}-\dfrac{\omega_{c}}{\omega}\right).
	\label{persi01}
	\end{equation}
Thus, the total current $I$ correspondent to the contribution of the all $N_{e}$ electrons will be given by 
	\begin{equation}
	I=\sum^{N_{e}}_{n,m}I_{n,m}\qquad E_{n,m}\le \epsilon_{f},
	\label{persi02}
	\end{equation}
whose behavior with the magnetic field applied for different values of anisotropy parameter is presented in Fig. \ref{anispersi} for $l=0$. In Eq. (\ref{persi01}), we see that the existence of persistent current does not necessarily depend on the presence of a magnetic field if $l\neq 0$. 
	
The amplitudes are suppressed when $\omega_{c}$ grows as a diamagnetic response of the system to the penetration of $B$, which is indicated by the negative sign. This effect is more intense for the case with $\alpha=1$. When curvature arises, the anisotropic-oblate systems suffer a faster decay than the prolate and isotropic ones. We can also display a relation between the magnetization (\ref{mag02}) and persistent current (\ref{persi01}) using the effective radius of the states $r_{n,m}=\lambda\sqrt{2L}$ \cite{AdP.2019.531.1900254}:
	\begin{equation}
	\mathcal{M}_{n,m}=\pi(\alpha r_{n,m})^{2} I_{n,m} - \frac{e\hbar}{m_{t}}\left[\alpha^2\left(n+\dfrac{1}{2}\right) \frac{\omega_{c}}{\omega}\right],\label{persi03}
	\end{equation}
where we also identify a diamagnetic term which vanishes when $\omega_{c}<<\omega$. At zero magnetic field, the magnetization expression is similar to the well-known result for a classical ring with a circular current. However, there is an extra $\alpha$ term we can interpret as a global geometric effect due to the conical geometry of the surface besides its influence in $L$. This $\alpha$ term in Eq. (\ref{persi03}) explains the change in relative amplitudes between magnetization and persistent current for weak fields shown in Figs. \ref{anismagnetizationlow} and \ref{anispersilow}. In fact, opposite to magnetization, systems with more curvature present higher amplitudes of persistent current. 
	
Furthermore, we also analyze the numerical behaviors in the extreme regimes of $B$ shown in Figs. \ref{anispersilow} and \ref{persihigh}, where one can recover the aspects discussed in Section 7 about the state transitions per period and the period increasing to $\approx 0.001463\hspace{0.05cm}\text{T}$. 
	
Anisotropic-oblate systems present higher amplitudes in the weak magnetic field. For strong $B$, we see more clearly the effect of curvature on both amplitudes and the period of the AB oscillations. Since there is only one occupied subband in this regime, there is only one transition per period, producing a ``sawtooth'' behavior \cite{PRB.1999.60.5626, PEREIRA2021114760}. We can also identify a slight amplitude difference between the anisotropic cases, where it reduces as the $R$ parameter increases. The ``phase'' difference in the AB oscillations is due to the depopulation of the $n=1$ subband shown at the last peak of the dHvA oscillations in Fig. \ref{fermienergy}.
	
	\section{Effects in the orbital gap of spin qubits}\label{sec9}
	
The last subject we discuss in this work is the behavior of the orbital gap in the quantum ring when affected by curvature and anisotropy since this is an important quantity for the development of the so-called \textit{spin qubits}, which can play a key role for the development of quantum information and quantum computation that are currently areas of great interest in Physics. The \textit{qubit} or \textit{quantum bit} is the unity of information in a quantum computer.  
	
Spin qubits are two-level systems formed by a quantum superposition of states in the same orbital $\chi_{n,m}(r,\varphi)$ of the ring but with opposite spins $s=\pm 1/2$. The general state $\ket{Q_{\text{spin}}}$ of the system is
	\begin{equation}
	\ket{Q_{\text{spin}}}=a\ket{\chi_{n,m},+1/2}+b\ket{\chi_{n,m},-1/2},
	\label{obgap01}
	\end{equation}
where the complex coefficients $a$ and $b$ must satisfy the condition $|a|^2+|b|^2=1$. However, to make this system stable against thermal excitations and external fields fluctuation which would cause decoherence to the system, the relation $k_{B}T<<\Delta_{s}<<\Delta_{\text{orb}}$ must hold, where the orbital gap $\Delta_{\text{orb}}$ is the maximum energy difference between a given orbital and its first excited orbital (see Fig. \ref{fluxqubit}). The spin gap $\Delta_{s}$ is the energy splitting between the two opposite spin states correspondent to the same orbital and increases with the magnetic flux \cite{MSzopa2010}.
	
The ratio $\Delta_{s}/\Delta_{\text{orb}}$ is of great interest for the quantum computation since the more remarkable is $\Delta_{\text{orb}}$ related to $\Delta_{s}$, more stable is the system. The two-level system provided by the electron spin will acquire more stability against magnetic field fluctuations because the spin states become well separated from the other orbitals. Thus, we investigate the influence of the anisotropy parameter $R$ and curvature parameter $\alpha$ for this property in the quantum ring. The results are illustrated in Fig. \ref{gaporbital}.
The orbital gap presents higher values for the same curvature configuration as $R$ decreases. Moreover, for the same parameter $R$, as the system increases its curvature, the quantity $\Delta_{\text{orb}}$ also increases. Therefore, combining these two properties (anisotropy and curvature) can provide more stable materials for developing qubits. 
	
	\section{Conclusions}\label{sec10}
	
We have studied the electronic properties, magnetization, persistent current, and modifications in the orbital gap of a 2DEG provided by the effective mass anisotropy in a quantum ring with conical geometry. We analyze the system for a class of semiconductors with anisotropic effective mass parameters $R$ subjected to a uniform magnetic field. Effects of surface curvature and anisotropy mass revealed several physical implications for the electronic properties of the ring, such as the modification in the minimum of the electronic subbands when investigated under magnetic field changes and different values of the anisotropy parameter $R$. We have also argued that the Fermi energy of the system is increased as $\alpha$ is reduced, with consequent physical impact on the other quantities derived from it. The results of our analysis suggest that one could regulate the size of the rings by using materials with different $R$ and $\alpha$ parameters for the same number of carriers in the 2DEG, which would be a good prospect in the context of mesoscopic device fabrication. In contrast, some $\alpha$ values were identified because of the region where the derivative of the ring width curve represents a sharp change. This is closely related to the addition of another occupied subband, and consequently, another spike in the Fermi energy should arise.
	
In characterizing the magnetization, our study has shown that the curvature of the ring significantly affects the magnetization oscillations caused by subband depopulation, commonly referred to as dHvA oscillations. The presence of ring curvature leads to a reduction in the amplitude of these oscillations. Additionally, the anisotropy of the system also plays a role in influencing the amplitudes, with lower values observed as the $R$ parameter increases. In the regime of weak magnetic fields, magnetization oscillations arise from the intersection of energy states, resulting in AB oscillations. The impact of surface curvature and anisotropic mass is also present in electronic transport and significantly affects the persistent current amplitudes. We have verified this in weak and strong magnetic field regimes. 
	
For the investigation of the persistent currents, we show their amplitudes are strongly suppressed as the magnetic field increases. This decay rate can be reduced for systems with curvature, as pointed out in Fig.  \ref{persihigh}. Also, the anisotropic-oblate systems presented the higher low magnetic field amplitudes of persistent current but suffered a faster decay than the other cases.
	
Our work was concluded by investigating the effects of curvature and anisotropic mass on the orbital gap of spin qubits. Within this perspective, we have argued that the physical implications of these effects may have good benefits in the future, such as helping to improve techniques for fabricating more stable materials for developing spin qubits in quantum information and quantum computation. 
	
\section*{Acknowledgments}
	
This work was partially supported by the Brazilian agencies CAPES, CNPq, and FAPEMA. Edilberto O. Silva acknowledges the support from the grants CNPq/306308/2022-3, FAPEMA/UNIVERSAL-06395/22, FAPEMA/APP-12256/22. This study was financed in part by the Coordena\c{c}\~{a}o de Aperfei\c{c}oamento de Pessoal de N\'{\i}vel Superior - Brazil (CAPES) - Finance Code 001.

\bibliographystyle{apsrev4-2}

\end{document}